\documentclass[a4paper]{article}
\usepackage{jcappub} 
\usepackage[utf8]{inputenc}
\usepackage{amsmath,amssymb}
\usepackage{graphicx}
\usepackage{hyperref}
\usepackage{natbib}
\usepackage[normalem]{ulem}
\pdfoutput=1

\usepackage{todonotes}
\usepackage{multirow}
\author[a,1]{Sudipta Das\note{Corresponding author}}
\author[a]{Manisha Banerjee}
\author[b]{Nandan Roy}
\affiliation[a]{Department of Physics, Visva-Bharati, Santiniketan -731235, India.}
\affiliation[b]{The Institute for Fundamental Study~ ``The Tah Poe Academia Institute", Naresuan University, Phitsanulok 65000, Thailand}
\emailAdd{sudipta.das@visva-bharati.ac.in}
\emailAdd{banerjee.manisha717@gmail.com}
\emailAdd{royn@nu.ac.th}

\title{Dynamical System Analysis for Steep Potentials}
 
\abstract{In this work we have performed the dynamical system analysis for steep(er) exponential potentials considering different values of the steepness index $n$. We have performed the analysis using centre manifold theory as well as by employing numerical method. We have shown that in most of the cases, the higher values of steepness index corresponds to an unstable solution. We have shown that with this steep(er) potentials one can not have a phase transition from dark matter to dark energy in the past.}

\begin{document} 

\maketitle
\section{Introduction}
Over the last decade one of the primary objectives of the cosmologists is to explain the reason behind the observed late time acceleration of the universe \cite{riess, perlmutter, tonry}. 
The true nature of dark energy (DE), the driver of this late time acceleration, is still unknown to us. A number of cosmological models have been explored to understand the true nature of DE \cite{ varun, Copeland, peebles}. From theoretical perspective, it is very hard to conclude whether the dark energy component is a cosmological constant or a dynamical quintessence field and the search is still on for a suitable candidate for DE. 
\par A quintessence DE model is usually characterized by a scalar field which rolls down a potential such that it provides an effective negative pressure and generates acceleration. A wide variety of quintessence potentials exist in literature which serve as a good candidate for dark energy. For an extensive review on quintessence potentials, one can look into \cite{Copeland, sahni}. 
\par Recently, a new form of quintessence potential, viz. steep exponential potential, have gained interest where the potential has the form \cite{wali}
\begin{equation}\label{steep}
V(\phi) = V_0 e^{\alpha(\frac{\phi}{M_p})^n},
\end{equation}
$V_{0}$ and $\lambda$ are model parameters. For $n=1$, equation (\ref{steep}) represents an exponential potential which has been studied extensively both in the context of inflation as well as late time acceleration \cite{sahnits, copelandliddle}. However, for $n>1$, the potential is steeper than the standard exponential potential. The parameter $n$ is a measure of the steepness of the potential. This type of steep exponential potentials have been explored to explain inflation as well as late time cosmic acceleration at one go \cite{wali}. Recently Sahalam et al. \cite{shahalam} have shown that steeper exponential potentials with $n=2$ and $n=3$ are consistent with the observational data and can very well depict the late time evolution picture of the universe. 
\par In the present work, we try to look into these steeper potentials from a different outlook. As already mentioned, the steeper exponential potentials are found to be consistent with late time acceleration. We are interested to know whether this kind of potentials can depict the evolution history of the universe consistently or not. To be more precise, whether with a steep(er) potential one can obtain a DE dominated phase preceded by a matter dominated era so as to explain structure formation of the universe. We have used the dynamical systems analysis to study the cosmological behaviour of these potentials. Dynamical systems analysis is a very powerful and elegant tool to study qualitative behaviour of nonlinear systems without studying it quantitatively. This approach has already been effectively used to study the true dynamical nature of a number of cosmological models (for details see the review \cite{dynamicalreview} and the references therein). This tool allows us to perform a preliminary analysis of any particular theoretical model suggesting which 
deserve further attention or which can not provide a consistent model. Keeping this in mind, in the present work we perform dynamical system analysis for steep(er) potentials for different values of $n$, namely, $n = 2, 3$ and $4$ in order to check the viability of the same. To do so, we first express the system of equations in autonomous form and find out the critical points for the system. The critical points turn out to be non-hyperbolic in nature. As linear stability theory can not be applied for non-hyperbolic critical points, we perform the analysis using centre manifold theory (CMT) as well as through perturbative approach and try to analyse the stability of systems characterized by steeper exponential potentials. 
\par The organization of the paper is as follows. In section 2, we provide the basic framework for the model and in section 3 we look into the stability of the system  using CMT. Section 4 deals with the analysis in a perturbative approach and in the final sections, we summarize the results.

\section{Scalar Field Dynamics with Steep Potential: Basic Framework }
For a spatially flat ($k=0$) Friedmann-Robertson-Walker (FRW) universe given by 
\begin{equation}
ds^2 = dt^2 - a^2(t)[dr^2 + r^2 d\theta^2 + r^2 sin^2\theta d\phi^2],
\end{equation}
the Einsteins field equations and the conservation equations for a canonical scalar field model are  given as ($p_m=0$), 
\begin{equation}\label{fe1}
3H^{2}=\rho_{m} + \frac{1}{2}{\dot{\phi}}^{2} + V(\phi) 
\end{equation}
\begin{equation}\label{fe2}
2{\dot{H}} + 3H^{2}=-\frac{1}{2}{\dot{\phi}}^{2} + V(\phi) 
\end{equation}
\begin{equation}\label{ce1}
{\dot{\rho}}_{\phi} + 3H(\rho_{\phi} + p_{\phi})=0  
\end{equation}
\begin{equation}\label{ce2}
{\dot{\rho}}_{m} + 3H\rho_{m}=0 
\end{equation}
where $H = \frac{\dot{a}}{a}$ is the Hubble parameter, $V(\phi)$ is the potential for the scalar field $\phi(t)$. We have chosen to work in natural units in which $8\pi G = c = 1$. \\ 
 As mentioned earlier, recently studies are being carried out considering a steep(er) exponential potential  where the form of the potential is considered to be $V(\phi) = V_0 e^{\alpha(\frac{\phi}{M_p})^n}$ \cite{shahalam}. The power $n$ determines the steepness of the potential. It has been shown that this simple extension of an exponential potential (the power $n \ne 1$ makes this different from an exponential potential) allows to capture late-time cosmic acceleration and is capable of retaining the tracker behavior \cite{shahalam}. In what follows, we will try to look into the stability of such scalar field dark energy models having a very steep potential.  
\subsection{Dynamical System Study}
In this section we rewrite the Einstein field equations (\ref{fe1}) - (\ref{ce2}) for this model as {\bf autonomous} system and study the stability of the critical points.
We define the dimensionless parameters 
\begin{equation}\label{param}
x=\frac{\dot{\phi}}{\sqrt{6}H}, ~y=\frac{\sqrt{V}}{\sqrt{3}H}, ~\lambda=\frac{1}{V}\frac{d V}{d\phi}, ~\Gamma=\frac{V V_{\phi\phi}}{V_{\phi}^2}
\end{equation}
where $V_{\phi} = \frac{dV}{d\phi}$ and $V_{\phi\phi} = \frac{d^2V}{d\phi^2}$.\\
In terms of these new variables, the system of equations can be rewritten as an autonomous system as
\begin{equation}\label{xprime}
x'=\frac{1}{2}\lbrace3x^3-3x[y^2+1]-\sqrt{6}y^2\lambda\rbrace
\end{equation}
\begin{equation}\label{yprime}
y'=-\frac{1}{2}y\lbrace-3x^2+3(y^2-1)-\sqrt{6}x\lambda\rbrace
\end{equation}
\begin{equation}\label{lambdaeqn}
\lambda'=\sqrt{6}\lambda^2[\Gamma - 1] x
\end{equation}
Here a prime indicates derivative with respect to $N=ln a$. \\
From equation (\ref{param}), one can at once define the effective equation of state parameter as well as the equation of state parameter for the scalar field as 
\begin{eqnarray}\label{wphi}
w_{eff} = \frac{p_{\phi}}{\rho_{\phi} + \rho_{m}} = x^2 - y^2, ~w_{\phi} = \frac{p_{\phi}}{\rho_{\phi}} = \frac{x^2 - y^2}{x^2 + y^2}\\ \nonumber
\Omega_{\phi} =x^2 + y^2 \mathrm{~~and~~} \Omega_m = 1-x^2 - y^2
\end{eqnarray}
For a simple exponential potential $V(\phi) = V_0 e^{\alpha(\frac{\phi}{M_p})}$, one immediately gets $\Gamma = 1$. However, for the particular choice of steep potentials given in equation (\ref{steep}), 
\begin{equation}\label{gamma}
\Gamma=1+(n-1)(n\alpha)^\frac{1}{n-1}\frac{1}{\lambda^\frac{n}{n-1}}
\end{equation}
For simplicity, we have chosen $M_p = 1$. \\
Equation (\ref{lambdaeqn}) can then be rewritten as (following \cite{dynamicalreview} for the power law case)
\begin{equation}\label{lambdaprime}
\lambda'=\sqrt{6}\lambda^2[\Gamma(\lambda) - 1] x = \sqrt{6}\lambda^{\frac{n-2}{n-1}}(n-1)(n\alpha)^\frac{1}{n-1}x
\end{equation}
To find the stability of the system we need to find out the critical points of the system which are obtained from the simultaneous solutions of the equations $x^{'}=0$, $y^{'}=0$ and $\lambda^{'}=0$.\\
The stability  of a fixed point can be determined from the eigenvalues of the Jacobian matrix constructed to linearise the system around it.
\begin{equation}
  J=\begin{pmatrix}
\frac{dx'}{dx} & \frac{dx'}{dy}  & \frac{dx'}{d\lambda} \\\\
\frac{dy'}{dx}  & \frac{dy'}{dy}  & \frac{dy'}{d\lambda} \\\\
\frac{d\lambda'}{dx}  & \frac{d\lambda'}{dy}  & \frac{d\lambda'}{d\lambda} \\
\end{pmatrix}
\end{equation}\\
If the real parts of the eigenvalues are not zero then the fixed point is called hyperbolic fixed point and one can use the linear stability method to find out stability of it. If the real part of any eigenvalue is zero then the fixed point is non-hyperbolic in nature. For a non-hyperbolic fixed point one can not use linear stability analysis. One needs to use other analytical tools like Lyapunov function or Central Manifold Theorem or one can use different numerical methods to study it.\\
 For the system of equations given by (\ref{xprime}), (\ref{yprime}) and (\ref{lambdaprime}), the last term of the Jacobian matrix comes out as
\begin{equation}\label{jacobian}
\frac{d\lambda'}{d\lambda} = \sqrt{6} (n-2) (n\alpha)^\frac{1}{n-1} \frac{1}{{\lambda}^{1/(n-1)}}x
\end{equation}
\begin{itemize}
\item For $n = 1, ~\Gamma = 1$ (by equation (\ref{gamma})) and $\lambda' = 0$ always and $\frac{d\lambda'}{d\lambda}$ does not exist. \\
\item For $n = 2, ~\lambda'= 2\sqrt{6} \alpha x$ which implies $\frac{d\lambda'}{d\lambda} = 0$. \\
In that case, $\lambda'= 0$ indicates that either $\alpha = 0$ or $x = 0$.    $\alpha = 0$ implies that $V(\phi)$ = constant which does not characterize an exponential potential and is thus uninteresting. On the other hand, $x = 0$ implies either $\phi$ is a static field or $H \rightarrow \infty$. This condition eventually refers to the very early phase of evolution.  
\item For $n = 3, ~\lambda'= 2 \sqrt{6} ~{(3\alpha)}^{1/2} {\lambda}^{1/2}x$.\\
In this case, $\lambda'= 0 \rightarrow$ indicates that either $\alpha = 0$ or $x = 0$ (which have been discussed earlier) or $\lambda = 0$.\\ 
From equation (\ref{jacobian}), it is evident that $\lambda = 0$ will blow up the last term of the Jacobian. This is indeed the case for all higher values of $n$ as well.
\end{itemize}

\par To {\bf overcome} this problem, we make a change of variable as follows. We replace 
\begin{equation}
z=\frac{(\lambda)^{\frac{1}{n-1}}}{1 + (\lambda)^{\frac{1}{n-1}}} \Rightarrow \lambda = (\frac{z}{1-z})^{(n-1)}\\
\end{equation}
When $\lambda = 0$ we get $z=0$ and in the limit $\lambda \rightarrow \infty$, we have $z = 1$. This means that the new variable $z$ is bounded as $0 \le z \le 1$. A similar approach has been used to study the phase space of power-law system in \cite{ng, dynamicalreview}. 

In terms of this new variable, equations (\ref{xprime}), (\ref{yprime}) and (\ref{lambdaprime}) comes out as 
\begin{eqnarray}
x'&=&\frac{1}{2}\lbrace3x^3-3x[y^2+1]-\sqrt{6}y^2(\frac{z}{1-z})^{(n-1)}\rbrace, \label{xprimez} \\
y'&=&-\frac{1}{2}y\lbrace-3x^2+3(y^2-1)-\sqrt{6}x(\frac{z}{1-z})^{(n-1)}\rbrace, \label{yprimez} \\
z'&=&\sqrt{6}x(n\alpha)^\frac{1}{n-1}(1-z)^2. \label{zprimez} 
\end{eqnarray}

Again as evident from the above equations, the last term in both (\ref{xprimez}) and (\ref{yprimez}) diverges as $z\rightarrow 1$. \\
In order to remove these infinities we multiply the right hand sides of (\ref{xprimez}) -(\ref{zprimez}) by $(1-z)^{(n-1)}$. The idea is as follows :\\
For any dynamical system $x' = f(x)$, the stability properties are determined by putting $x'= 0$. One can thus construct a new dynamical system as $x' = \zeta(x)f(x)$ for any positive defined function $\zeta(x) > 0$ without any loss of generality. This will indeed provide the same critical points with same stability properties. This approach has been used in \cite{dynamicalreview} for the power law case. \\
In this case also we have multiplied the right hand sides of (\ref{xprimez}) -(\ref{zprimez}) by $(1-z)^{(n-1)}$ which is positive definite as $0 \le z \le 1$. So effectively the nature of critical points and the stability properties remain the same. We are just removing the divergent terms keeping the rest of the phase plane invariant. \\
With this we obtain

\begin{eqnarray}
x^{'}&=&\frac{1}{2}(1-z)^{(n-1)}\lbrace3x^3-3x[y^2+1]\rbrace-\sqrt{\frac{3}{2}}y^2z^{(n-1)}, \label{xprimezfinal}\\
y^{'}&=&-\frac{1}{2}y(1-z)^{(n-1)}\lbrace-3x^2+3(y^2-1)\rbrace+\sqrt{\frac{3}{2}}xy z^{(n-1)}, \label{yprimezfinal}\\
z^{'}&=&\sqrt{6}x(n\alpha)^\frac{1}{n-1}(1-z)^{(n+1)}, \label{zprimezfinal}
\end{eqnarray}

The above set of equations are now regular at $z=1$. 
\subsection{Critical points and their analysis}
We next consider the critical points for the dynamical system given by equations (\ref{xprimezfinal}) -(\ref{zprimezfinal}). The critical points and the corresponding eigen values are listed in Table \ref{criticalpoint}.

\begin{table}[!h]
\begin{center}

\begin{tabular}{|c|c|c|c|c|c|}
\hline
Critical Points &x &y &z & \multicolumn{2}{c|}{Eigen values}\\

\hline
A &0 &0 &Any & \multicolumn{2}{c|}{0,~$-\frac{3}{2}(1-z)^{(n-1)}, ~\frac{3}{2}(1-z)^{(n-1)}$}\\
\hline
B &Any &0 &1 &\multicolumn{2}{c|}{$0,~0,~\sqrt{\frac{3}{2}} x$}\\
\hline
\multirow{2}{*}{C} &\multirow{2}{*}{0} &\multirow{2}{*}{1}  &\multirow{2}{*}{0}& for~$ n=2 $ & -3,$\frac{1}{2}(-3 \pm \sqrt{9-24\alpha})$ \\
\cline{5-6}
&&&& for~$ n=3,4,5...$ &-3,-3,0 \\
\hline
\multirow{2}{*}{D} &\multirow{2}{*}{0} &\multirow{2}{*}{-1}  &\multirow{2}{*}{0}& for~$ n=2 $ & -3,$\frac{1}{2}(-3 \pm \sqrt{9-24\alpha})$\\
\cline{5-6}
&&&&for~$ n=3,4,5...$ &-3,-3,0\\
\hline
\end{tabular}
\caption{Critical points of the system and the corresponding eigenvalues for different values of $n$.}
\label{criticalpoint}
\end{center}
\end{table}
As evident from Table \ref{criticalpoint}, all the critical points (except points C and D with $n=2$) have at least one zero eigen value which means that the points are non-hyperbolic and one can not use the linear  stability theory. In that case one can apply either centre manifold theory (CMT) or Lyapunov's method to analyse the nature of the non hyperbolic critical points analytically. One can also take help of the numerical programming to visualize the trajectories near these fixed point and study their stability. For a system which is higher than 2D, drawing a phase plot is not always enough to draw the conclusion about the stability. For example, figure 1 shows the 3 dimensional phase plots for the system for $n=2$, $n=3$ and $n=4$. For $n=2$ and $n=3$, the values of $\alpha$ have been chosen as $\alpha=8.87$ and $\alpha=3.7$ respectively which has been taken from the best fit values obtained by Shahalam et al. \cite{shahalam}.  While for $n=4$, we have chosen $\alpha=1.4$ as it seems that the best fit value of $\alpha$ decreases with increase in $n$. In Fig.1 the range of the $x, y$ coordinates is from $-1$ to $+1$ as these variables  are constrained by the Friedmann constrain equation $0 \le \Omega_{\phi} = x^2 + y^2 \le 1$. The $z$ axis varies from $0$ to $+1$ which comes from our mathematical construction. In the phase plot one can see the existence of the Friedmann circles on the $x-y$ plane.\\ 

\begin{figure}[ht]
\begin{center}
\includegraphics[width=0.32\columnwidth]{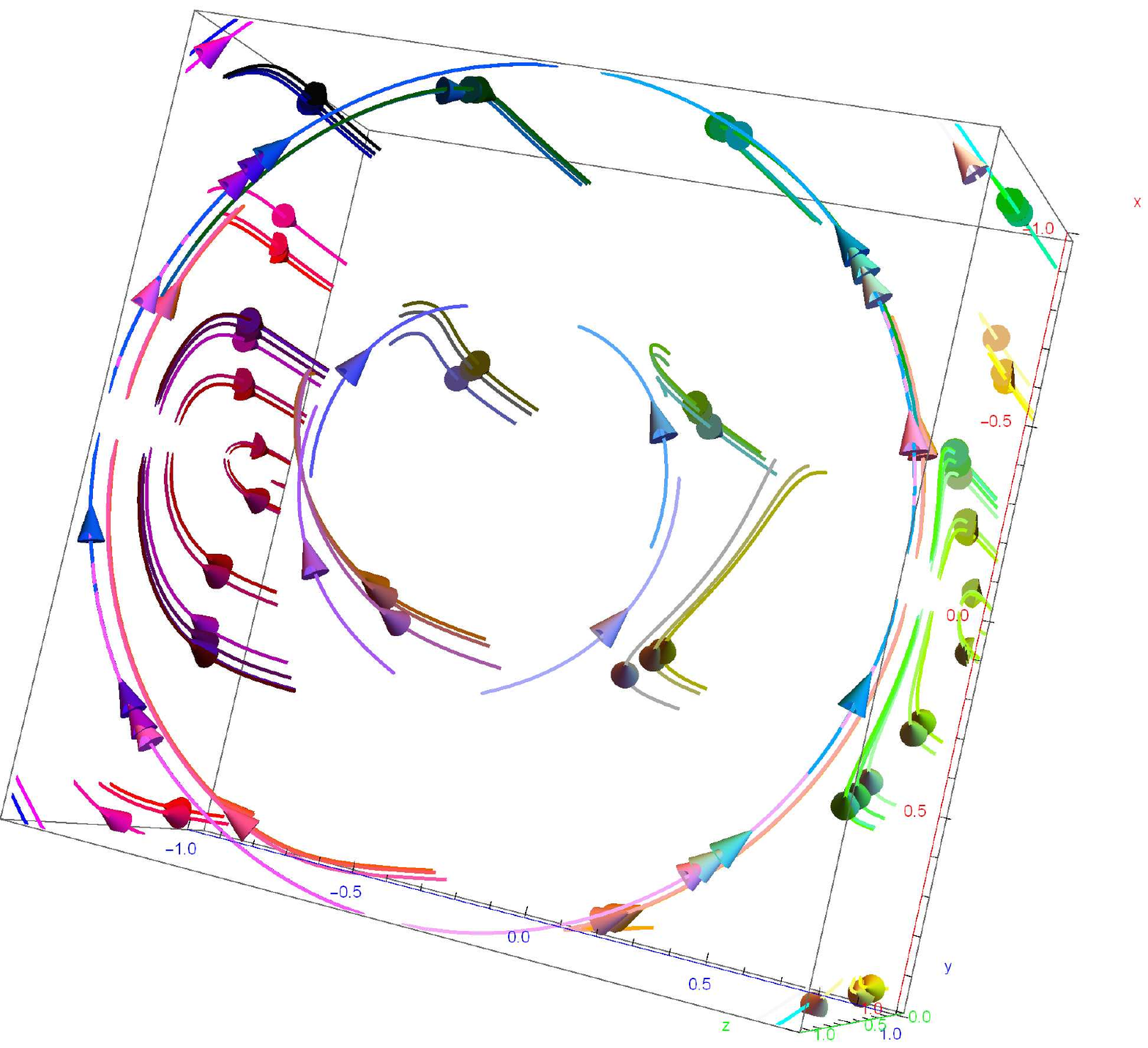}
\includegraphics[width=0.32\columnwidth]{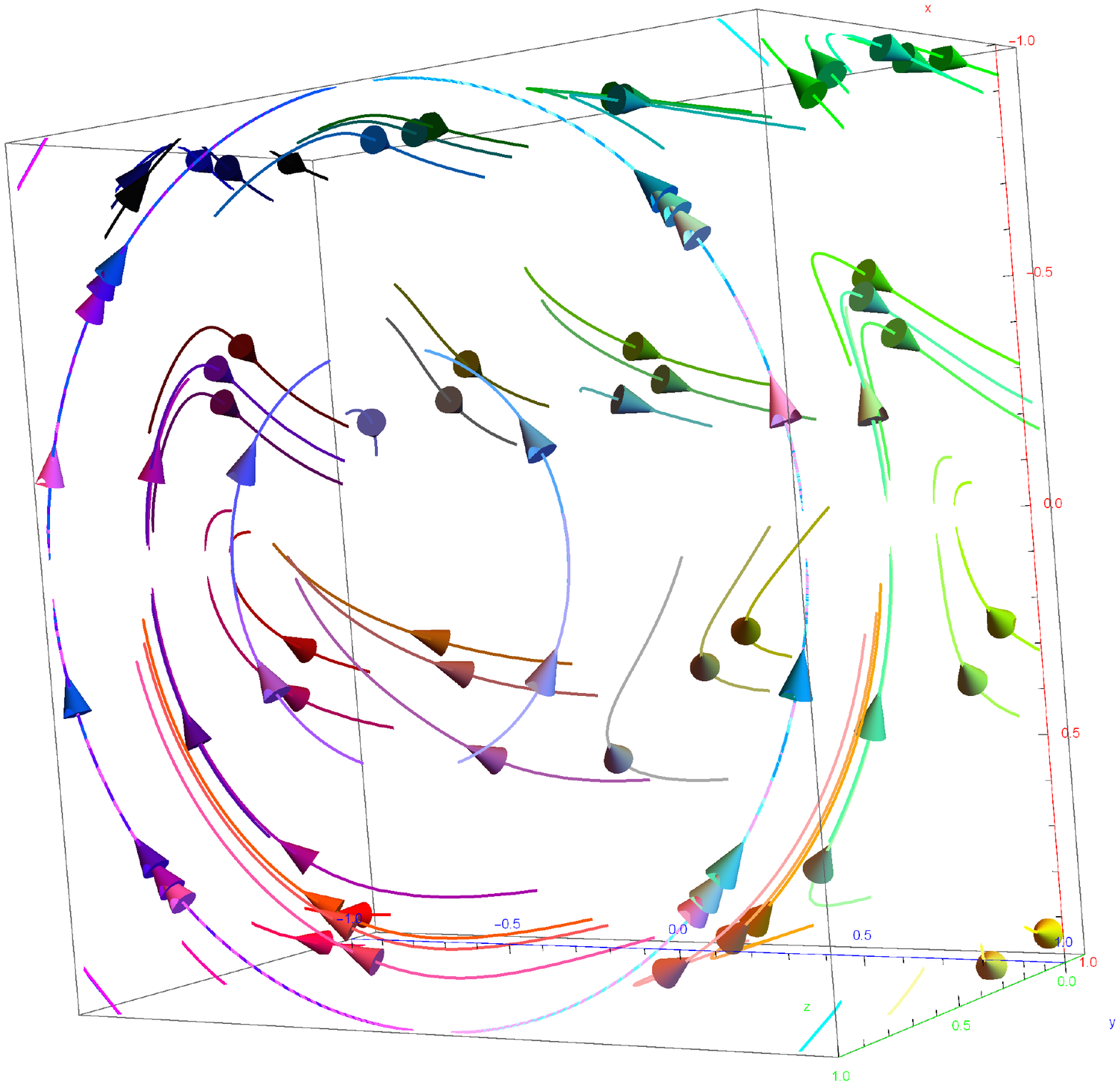}
\includegraphics[width=0.32\columnwidth]{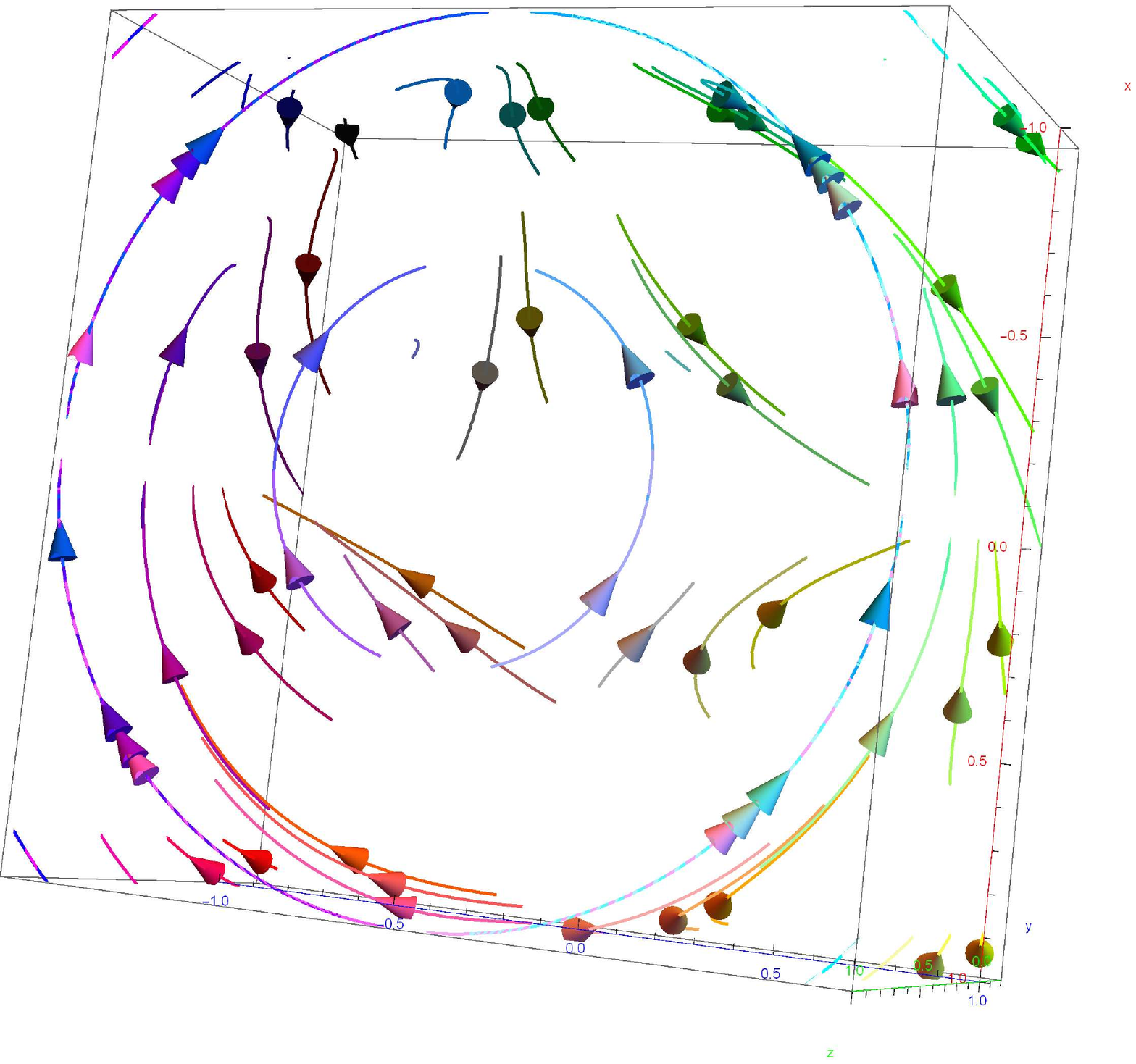}
\caption{\em 3 dimensional phase plots for $n=2, 3, 4$ are shown for $\alpha = 8.87, ~3.7$ and $1.4$ respectively.In all these plots the red line represents $x$-axis, blue line represents $y$-axis and the green line represents $z$-axis. }
\label{fig3D}
\end{center}
\end{figure}

From the 3D plots, it is very difficult to understand or analyze the nature of the critical points. As evident from Table \ref{criticalpoint}, points A have one zero eigen value and two non-zero eigen values given by $\pm \frac{3}{2}(1-z)^{(n-1)}$, such that they always bear equal and opposite values in the range $0\le z \le 1$.  This makes these points a saddle point. This can as well be visualized by looking at the centre of the 3D box in figure \ref{fig3D} which shows this saddle point nature. But for rest of the points no such inference can be drawn from figure \ref{fig3D}. Instead of phase plot analysis, one can study the stability of the non-hyperbolic fixed points from the behaviour of the perturbations around these fixed points (see e.g \cite{nandan1, nandan2, nandan3, zonu} for details). Depending on the time evolution of the perturbations it is very easy to draw the conclusion about the stability of a fixed point. \\

In this work in order to understand the true nature of the critical points, we  have used both centre manifold theory (CMT) and perturbative approach and compared the results obtainted by these two different methods of stability analysis. 
\section{Analysis using Centre Manifold Theorem}\label{CMT}

At the beginning of this section we provide a brief description of the basic principles of center manifold theory. For detailed analysis and mathematical background of center manifold theory, one can look into \cite{carr} - \cite{boehmer}. \\
As mentioned earlier, if the Jacobian matrix at the critical point contains zero eigen values, then the linear stability theory fails to provide information regarding the stability of that point. In centre manifold theory, the dimensionality of the system near that critical point is reduced  which helps to investigate the stability of the reduced system. The nature of stability of the reduced system depicts the same for the actual system at that point.\\
\par If the Jacobian matrix $J$ of a non-hyperbolic critical point  contains at least one eigenvalue having positive real part, then the corresponding critical point is always unstable, irrespective of whether it is hyperbolic or non-hyperbolic. Otherwise if the non-vanishing eigenvalues of $J$ (corresponding to non-hyberbolic critical points) contain all negative real parts, then  centre manifold is applied to find the stability. The CMT analysis involves the following steps :\\\\
1. At the beginning  the co-ordinates of the non-hyperbolic critical points  are shifted to the origin through suitable co-ordinate translation which leads to a set of autonomous equations in the new coordinates. \\\\
2. The new dynamical system is then rewritten in the following standard form

           \begin{align}
               u'&=A u+f(u,v)\label{cmteqn1} \\
                v'&=B v+g(u,v)\label{cmteqn2}
            \end{align}
                    where $(u,v)\in \mathbb{R}^c \times \mathbb{R}^s$ with $f$ and $g$ satisfying
                     \begin{align}
                          f(0,0)=0,\quad Df(0,0)=0\\
                          g(0,0)=0,\quad Dg(0,0)=0
                       \end{align}
Here we can see the critical point is located at the origin and $Df$ denotes the matrix of first derivatives of the vector valued function $f$. $A$ is a $c\times c$ matrix whose eigenvalues have zero real part, $B$ is $s\times s$ matrix whose eigenvalues have negative real part.\\
3. Next a function $h(u)$ is determined, which is usually approximated by a series expansion having at least $u^2$ term, such that $h(u)$ satisfies the following quasilinear partial differential equation
 \begin{align}\label{quasi}
 \mathcal{N}h(u)\equiv Dh(u)\left(A u+f(u,h(u))-B h(u)-g(u,h(u))\right)=0,
\end{align}
with $h(0)=0$ and $Dh(0)=0$.\\

4. By substituting the approximated solution of $h(u)$ obtained from the last equation, one can find the dynamics of the original system restricted to the center manifold as 
 \begin{equation}\label{cmtreduce}
 u'=A u+f(u,h(u))
 \end{equation}
 
 5. Equation (\ref{cmtreduce}) then takes the final form  $u'= k u^n$ where $k$ is a constant and  $n$ is a positive integer (the lowest order in the expansion). \\
 \begin{itemize}
     \item If $k < 0$ and $n$ is odd-parity, this corresponds to a stable system. This in turn implies the stability of the original system. \\
     \item In all other cases, the reduced system and hence the original system will lead to instability \cite{Perko}.
 \end{itemize}
     In what following subsections we explicitly show the CMT analysis for the critical points obtained in Table \ref{criticalpoint} in accordance to the steps outlined above. 
\subsection{Analysis of the critical points using CMT}
\subsubsection{\bf{Point A ($0, 0$, any)}} 
All the points with $x = y = 0$ are critical points for any value of $z$. This implies that the $z$-axis is a critical line. The $x = y=0$ values makes the effective equation of state $w_{eff}$ zero and leaves $w_{\phi}$ undetermined and one can not draw any inference regarding the accelerating solution of the system. However, as discussed earlier, the other two eigen values for points A are $\pm \frac{3}{2}(1-z)^{(n-1)}$. Since by definition $z$ is bounded as $0\le z \le 1$, these two eigen values will always bear opposite signatures which makes these points a saddle point.\\

    \subsubsection{ \bf Point B ($x$, $0, 1$)} 
    \par The eigen values for point B are ($0, 0, \sqrt{\frac{3}{2}} x$). Two of the eigenvalues are zero making these points non-hyperbolic. One can thus use the centre manifold theory to analyse the nature of the critical points. According to the centre manifold theory, the stable space is spanned by the eigen vectors of Jacobian matrix associated with eigen values having negative real part whereas the unstable space is spanned by the eigen vectors of Jacobian matrix associated to eigenvalues with positive real part. If the Jacobian has at least one eigen value with positive real part, then the corresponding critical point can never be stable, irrespective of it being hyperbolic or non-hyperbolic \cite{dynamicalreview}. Thus for points B there are two possibilities: \\
    (i) If $x>0$, one of the eigenvalues (viz. $\sqrt{\frac{3}{2}} x$) will be positive and thus will lead to an unstable configuration having a non-accelerating feature. So this fixed point may refer to the initial point in the phase space from which the universe started to evolve. \\  
    (ii) If $x < 0$, the eigen value mentioned above  will be negative and thus one should analyse the system using CMT.\\ 
    
    The details of centre manifold dynamics calculations for point B is given in appendix \ref{appendixA1}. Here we will summarize the results obtained by CMT analysis. For ease of calculation, let us rename the point B as ($\pm a$, 0, 1) where  $0 \leq a \leq 1$ because $x$ is restricted by Friedmann constraint equation as $-1 \leq x \leq 1$.  \\
To proceed we have to first transform the system of equations (\ref{xprimezfinal})-(\ref{zprimezfinal}) into the standard form by rescaling the co-ordinates such that
\begin{equation}
X=x\mp a
\end{equation}
\begin{equation}
Y=y
\end{equation}
\begin{equation}
Z=z-1
\end{equation}
This rescaling moves the point (${\pm a},0, 1$) to the origin ($0, 0, 0$) of the phase space.
 
Then equations  (\ref{xprimezfinal})-(\ref{zprimezfinal}) become
\begin{equation}\label{XprimeB}
    X'=\frac{1}{2}(-Z)^{n-1} \{3(X\pm a)^3-3(X\pm a)(Y^2+1)\}-\sqrt{\frac{3}{2}}Y^2(Z+1)^{n-1}
\end{equation}
\begin{equation}\label{YprimeB}
    Y'=-\frac{1}{2}Y(-Z)^{n-1} \{-3(X\pm a)^2+3(Y^2-1)\}+\sqrt{\frac{3}{2}}(X \pm a)Y(Z+1)^{n-1}
\end{equation}
\begin{equation}\label{ZprimeB}
    Z'=\sqrt{6}(X \pm a)(n\alpha)^\frac{1}{n-1} (-Z)^{n+1}
\end{equation}
The Jacobian matrix of this system of equations evaluated at $(0,0,0)$ is obtained as  
  \begin{equation}
   J_3=    \begin{pmatrix}
      0~~ & 0~~ & -\frac{3}{2}\lbrace(\pm a)^3-(\pm a)\rbrace\\
       0~~ & \pm \sqrt{\frac{3}{2}}a~~ &0 \\
       0~~ & 0~~ & 0
    \end{pmatrix}      ~~~~~\mathrm{for}~~  n=2, ~~a \ne 0,1
  \end{equation}
 
\begin{equation}
   J_4=    \begin{pmatrix}
      0~~ & 0~~ & 0\\
       0~~ & \pm \sqrt{\frac{3}{2}}a~~ &0 \\
       0~~ & 0~~ & 0
    \end{pmatrix}      ~~~~~\mathrm{for}~~ \begin{cases} n=3,4,5..., ~~a = \mathrm{any} \\
                                            n=2, ~~a = 0,1
                                            \end{cases}
  \end{equation}
    
\begin{enumerate} 
 \item{$n=2, ~a=0$ :}\\
 In this case, the Jacobian $J_4$ becomes a null matrix and thus one cannot apply CMT theory to analyse the nature of critical points. So for $n=2, ~a=0$, CMT theory fails.\\
 
 \item {For $n=2, ~a=1$ :}
 The system of equations are brought into the standard form by introducing a new set of variables $u$, $v$, $w$ as 
 $$X=u$$
 $$Y=w$$
 $$Z=v$$
 With this substitution above set of equations (\ref{XprimeB}) - (\ref{ZprimeB}) take the final form as (detailed calculation has been provided in appendix \ref{appendixA2})
    \begin{equation}
  \begin{pmatrix}

  u'\\
  v'\\
  w'\\
    \end{pmatrix}  
    =
    \begin{pmatrix}
   0 & 0& 0\\
    
    0 &0&0\\
    0& 0& \pm \sqrt{\frac{3}{2}}&\\

    \end{pmatrix}
    \begin{pmatrix}
  u\\
  v\\
  w\\
    \end{pmatrix}  
   +
\begin{pmatrix}
  g_1\\
  g_2\\
  f\\
    \end{pmatrix} 
    \end{equation}
    where $g_1$, $g_2$ and $f$ are polynomials in $(u, v, w)$ greater than $2$. 
Following the method described in section \ref{CMT}, we arrive at the dynamics of the system restricted to the centre manifold given by (for detailed calculation refer to appendix \ref{appendixA2})
\begin{equation}\label{aDotw}
    \Dot{w}=Aw + F(w, h(w)) = \pm \sqrt{\frac{3}{2}}w-\frac{1}{2}\sqrt{\frac{3}{2}}w^3+O(w^4)
\end{equation}
At the lowest order of equation (\ref{aDotw}) we have obtained an odd-parity term with coefficient $\pm \sqrt{\frac{3}{2}}$. According to CMT theory, if the lowest order is odd-parity term with positive co-efficient then the system is unstable and if the co-efficient of the lowest order odd-parity term is negative, then the system is stable. \\
Thus for $n=2$ point  $B$ will be stable for negative value of $a$ and will be unstable for positive value of $a$. This also matches with the discussion provided at the beginning of section $3.1.2$ where it has been mentioned that for point $B$, if $x$ (or equivalently $a$) is positive, then the system will always be unstable.   

\item {For $n=3$, $a = any$ }\label{appendixA3}

 For $n=3$ also, the system of equations are not in the required standard form and thus we make a change of variable as 
 $$X=u$$
 $$Y=w$$
 $$Z=v$$
 With this substitution above set of equations (\ref{XprimeB}) - (\ref{ZprimeB}) take the final form as
 \begin{equation}
  \begin{pmatrix}

  u'\\
  v'\\
  w'\\
    \end{pmatrix}  
    =
    \begin{pmatrix}
   0 & 0& 0\\
    
    0 &0&0\\
    0& 0& \pm \sqrt{\frac{3}{2}} a&\\

    \end{pmatrix}
    \begin{pmatrix}
  u\\
  v\\
  w\\
    \end{pmatrix}  
   +
\begin{pmatrix}
  g_1\\
  g_2\\
  f\\
    \end{pmatrix} 
    \end{equation}
 where $g_1$, $g_2$ and $f$ are polynomials in $u$, $v$, $w$ of order higher than $2$. The detailed calculation has been provided in appendix \ref{appendixA3}. 
  
Following the standard procedure of CMT analysis, the dynamics of the system restricted to the centre manifold is obtained as
\begin{equation}\label{a2Dotw}
    \Dot{w}=Aw + F(w, h(w)) = \pm \sqrt{\frac{3}{2}}aw \mp \frac{1}{2a}\sqrt{\frac{3}{2}}w^3+O(w^4)
\end{equation}
At the lowest order of equation (\ref{a2Dotw}) we have obtained an odd-parity term. Hence, like before, point  B for $n=3$ will be stable or unstable depending on the sign of the critical point.  \\

\item{} A similar analysis has been carried out for $n=4, 5, ....$, $a = any$. It has been found that in all the cases, the lowest order appears to be an odd-parity term with co-efficient $\pm \sqrt{\frac{3}{2}}a$ and thus the stability will depend upon whether $x$ is positive or negative. 

\item {For $n=2$, $a \ne 0, ~1$ : }
In this case also, the Jacobian $J_3$ or equivalently the system of equations are not in the standard form. To bring the system of equations in the required standard form, as  mentioned in section \ref{CMT}, one has to first find out the eigen vectors corresponding to the Jacobian matrix. Then with this set of eigen vectors one constructs the matrix $S$, which is the matrix of eigen vectors and finds its inverse $S^{-1}$. However, for point B, $n=2$, $a \ne 0, ~1$ case, the matrix of eigen vectors, $S$, happened to be a singular matrix and thus its inverse could not be determined. For detailed calculation refer to appendix \ref{appendixA4}. So centre manifold theory fails here and one cannot arrive at any conclusion regarding the dynamics of the universe. \\
In the next section we employ the numerical perturbation method to find the stability of this fixed point.  \\
    \end{enumerate}
    \subsubsection{\bf Point C ($0, 1, 0$)} 
    \par The eigenvalues for point C are $[-3, ~\frac{1}{2}(-3 \pm \sqrt{9-24\alpha}) ]$ for $n=2$ and $[-3, 0, -3]$ for other values of $n$ i.e. for $n= 3$, $4$ and so on.  We will summarize the results here and the detailed calculations can be found in appendix \ref{appendixB}. \\ 
    \begin{itemize}
       \item For n=2 : As evident from Table \ref{criticalpoint}, for point C, the eigenvalues corresponding to $n=2$ has one negative real value and two complex conjugate values with negative real part. This will always correspond to a stable solution. 
       
        \item For other values of $n$ : With the following rescaling of the co-ordinates of the system  
\begin{equation}
X=x
\end{equation}
\begin{equation}
Y=y-1
\end{equation}
\begin{equation}
Z=z
\end{equation}
the critical point (0, 1, 0) is moved to the origin (0, 0, 0) of the
phase space. The system of equations (\ref{xprimezfinal})-(\ref{zprimezfinal}) take the form
\begin{equation}\label{xprimepointcc}
    X'=\frac{1}{2}(1-Z)^{n-1} \{3X^3-3X(Y^2+2Y+2)\}-\sqrt{\frac{3}{2}}(Y^2+2Y+1)Z^{n-1}
\end{equation}
\begin{equation}\label{yprimepointcc}
    Y'=-\frac{1}{2}(Y+1)(1-Z)^{n-1} \{-3X^2+3(Y^2+2Y)\}+\sqrt{\frac{3}{2}}X(Y+1)Z^{n-1}
\end{equation}
\begin{equation}\label{zprimepointcc}
    Z'=\sqrt{6}X(n\alpha)^\frac{1}{n-1} (1-Z)^{n+1}
\end{equation} 
The system of equations are then brought into the diagonal form by introducing a new set of variables $u, v, w$ as (details given in appendix \ref{appendixB})
\begin{equation}\label{criticalcc}
  \begin{pmatrix}
  u\\
  v\\
  w\\
    \end{pmatrix}  
    =
    \begin{pmatrix}
    0~~ & 1~~& 0\\
    
   - \sqrt{\frac{2}{3}}(n\alpha)^\frac{1}{n-1}~~ &0~~ &0\\
    \sqrt{\frac{2}{3}}(n\alpha)^\frac{1}{n-1}~~ &0~~ &1\\

    \end{pmatrix}
    \begin{pmatrix}
  X\\
  Y\\
  Z\\
    \end{pmatrix}  
    \end{equation}
    Next we summarize the results obtained for $n=3$ and $n=4$ respectively. \\
    \begin{enumerate}
     \item For $n=3$:  
     Equation (\ref{criticalcc}) becomes   
      \begin{equation}
       \begin{pmatrix} 
       u \\ v\\ w\\ 
        \end{pmatrix} =  
        \begin{pmatrix}
    ~0 & 1& 0\\
    - \sqrt{2\alpha} &0 &0\\
    ~\sqrt{2\alpha} &0 &1\\
    \end{pmatrix}
    \begin{pmatrix} X \\ Y\\ Z\\
     \end{pmatrix}
    \end{equation}
    Following the method of CMT analysis described in section \ref{CMT}, one obtains the dynamics of the system restricted to the centre manifold as (detailed calculations has been provided in appendix \ref{appendixB1})
\begin{equation}\label{cdotw3}
    \Dot{w}=Aw + F(w, h(w)) = -\sqrt{3\alpha}w^2+2(\sqrt{3\alpha} - \alpha)w^3+O(w^4)
\end{equation} 

The lowest order of equation (\ref{cdotw3}) has an even-parity term. According to CMT theory, an even power in the lowest order always leads to an unstable solution. Thus for $n=3$, point  C will always be unstable.  \\
\item For $n=4$ :
   The transformed system (refer to equation (\ref{criticalcc})) will take the form 
 \begin{equation}
  \begin{pmatrix}
  u\\
  v\\
  w\\
    \end{pmatrix}  
    =
    \begin{pmatrix}
    0~~ & 1~~& 0\\
    
   - \sqrt{\frac{2}{3}}(4\alpha)^\frac{1}{3}~~ &0~~ &0\\
  \sqrt{\frac{2}{3}}(4\alpha)^\frac{1}{3}~~ &0~~ &1\\

    \end{pmatrix}
    \begin{pmatrix}
  X\\
  Y\\
  Z\\
    \end{pmatrix}  
    \end{equation}
  which can also be written as 
  \begin{equation}
u = Y ~~\Rightarrow ~~Y = u
\end{equation}
\begin{equation}
v = - mX ~~\Rightarrow ~~X = -\frac{v}{m}
\end{equation}
\begin{equation}
w = mX + Z ~~\Rightarrow~~ Z = w + v
\end{equation}
where $ m=\sqrt{\frac{2}{3}}(4\alpha)^\frac{1}{3}$\\
 
Proceeding as before, the dynamics of the system restricted to the centre manifold is obtained as (see appendix \ref{appendixB2} for detailed calculations)
\begin{equation}\label{cdotwc1}
    \Dot{w}=Aw + F(w, h(w)) = -\sqrt{\frac{3}{2}}mw^3+O(w^4)
\end{equation}
At the lowest order of equation (\ref{cdotwc1}) we have obtained an odd-parity term with negative coefficient. Hence according to CMT theory, for $n=4$ point  C will always be stable.  
    \end{enumerate}
    \end{itemize}

    \subsubsection{\bf Point D ($0, -1, 0$) } 
    The eigen values for point D are same as that of point C :  
    $[-3, ~\frac{1}{2}(-3 \pm \sqrt{9-24\alpha})]$ for $n=2$ and $[-3, 0, -3]$ for higher values of $n$.
    \begin{itemize}
        \item For n=2 : As before, one of the eigen values corresponding to point D has negative real value and the other two have complex conjugate values with negative real part. This will always lead to a stable solution. 
        \item For other values of $n$ : As before, we transform the system of equations (\ref{xprimezfinal})-(\ref{zprimezfinal}) to the standard form by the following co-ordinate rescaling :  
\begin{equation}
X=x
\end{equation}
\begin{equation}
Y=y+1
\end{equation}
\begin{equation}
Z=z
\end{equation}
This shifts the critical point (0, -1, 0) to the origin (0, 0, 0) of the
phase space. Then equations (\ref{xprimezfinal})-(\ref{zprimezfinal}) becomes
\begin{equation}\label{xprimepointcc}
X'=\frac{1}{2}(1-Z)^{n-1} \{3X^3-3X(Y^2-2Y+2)\}-\sqrt{\frac{3}{2}}(Y^2-2Y+1)Z^{n-1}
 \end{equation}
\begin{equation}\label{yprimepointcc}
 Y'=-\frac{1}{2}(Y-1)(1-Z)^{n-1} \{-3X^2+3(Y^2-2Y)\}+\sqrt{\frac{3}{2}}X(Y-1)Z^{n-1}
\end{equation}
\begin{equation}\label{zprimepointcc}
     Z'=\sqrt{6}X(n\alpha)^\frac{1}{n-1} (1-Z)^{n+1}
\end{equation}
The system of equations are then expressed into the diagonal form by introducing a new set of variables given by (for detailed calculations refer to appendix \ref{appendixC})
\begin{equation}\label{criticaldc}
  \begin{pmatrix}
  u\\
  v\\
  w\\
    \end{pmatrix}  
    =
    \begin{pmatrix}
    0~~ & 1~~& 0\\
    
   - \sqrt{\frac{2}{3}}(n\alpha)^\frac{1}{n-1}~~ &0~~ &0\\
    \sqrt{\frac{2}{3}}(n\alpha)^\frac{1}{n-1}~~ &0~~ &1\\

    \end{pmatrix}
    \begin{pmatrix}
  X\\
  Y\\
  Z\\
    \end{pmatrix}  
    \end{equation}
 Next we will show the detailed calculations for two different values of $n$, i.e., $n=3$ and $n=4$.    
 \begin{enumerate} 
 \item{$n=3$ :}\\
For $n = 3$, the transformed system (\ref{criticaldc}) takes the form 
 \begin{equation}
  \begin{pmatrix}
  u\\
  v\\
  w\\
    \end{pmatrix}  
    =
    \begin{pmatrix}
    0~~ & 1~~& 0\\
    
   - \sqrt{2\alpha}~~ &0~~ &0\\
    \sqrt{2\alpha}~~ &0~~ &1\\

    \end{pmatrix}
    \begin{pmatrix}
  X\\
  Y\\
  Z\\
    \end{pmatrix}  
    \end{equation}
 Proceeding in a similar manner as above, we obtain the dynamics of the system as (Detailed calculation has been given in appendix \ref{appendixC1})
\begin{equation}\label{cdotwc3}
    \Dot{w}=Aw + F(w, h(w)) = -\sqrt{3\alpha}w^2+(6-2\sqrt{3}\alpha)w^3+O(w^4)
\end{equation}
At the lowest order of equation (\ref{cdotwc3}) we again obtain an even-parity term which indicates that for $n=3$ the dynamics of the system around point D will always be unstable.

\item {For $n=4$:}\\ 
The transformed system (\ref{criticaldc}) for $n=4$ takes the form 
 \begin{equation}
  \begin{pmatrix}
  u\\
  v\\
  w\\
    \end{pmatrix}  
    =
    \begin{pmatrix}
    0~~ & 1~~& 0\\
    
   - \sqrt{\frac{2}{3}}(4\alpha)^\frac{1}{3}~~ &0~~ &0\\
  \sqrt{\frac{2}{3}}(4\alpha)^\frac{1}{3}~~ &0~~ &1\\

    \end{pmatrix}
    \begin{pmatrix}
  X\\
  Y\\
  Z\\
    \end{pmatrix}  
    \end{equation}
  Again by following the same procedure, the final form of the dynamical system is obtained as (for details refer to appendix \ref{appendixC2})
\begin{equation}\label{cdotwc}
    \Dot{w}=Aw + F(w, h(w)) = -(4\alpha )^\frac{1}{3}w^3+O(w^4)
\end{equation}
At the lowest order of equation (\ref{cdotwc}) we have obtained an odd-parity term with negative coefficient which implies that for $n=4$ point D will always be stable. 
\end{enumerate}
\end{itemize}
\section{Numerical investigation of critical points} 

In this section we have used the numerical approach described before in section 2 to study the non-hyperbolic fixed points. This numerical method has been employed to study the stability of fixed points for a number of cosmological models \cite{nandan1, nandan2, nandan3, zonu}. This method is particularly interesting as one can study the stability of a critical point more easily as compared to the analytical method which need rigorous  mathematical calculations. In this method the system is perturbed in every direction near the particular fixed point in which we are interested and the perturbations are allowed to evolve with time (or equivalently with $N = ln a$).  If after the perturbations, the system comes back to the critical point as $N \rightarrow \infty$, then that fixed point is stable as it attracts the solutions/trajectories around it. If the system does not come back to the fixed point, which means the trajectories are repelled by the fixed point, then it is an unstable fixed point.

\par We have employed this perturbative approach for the critical points $B, C$ and $D$ listed in Table \ref{criticalpoint}. As point $A$ is always a saddle point, we need not employ any perturbative technique to test its stability criteria.  
\begin{figure}[ht]
\begin{center}
\includegraphics[width=0.32\columnwidth]{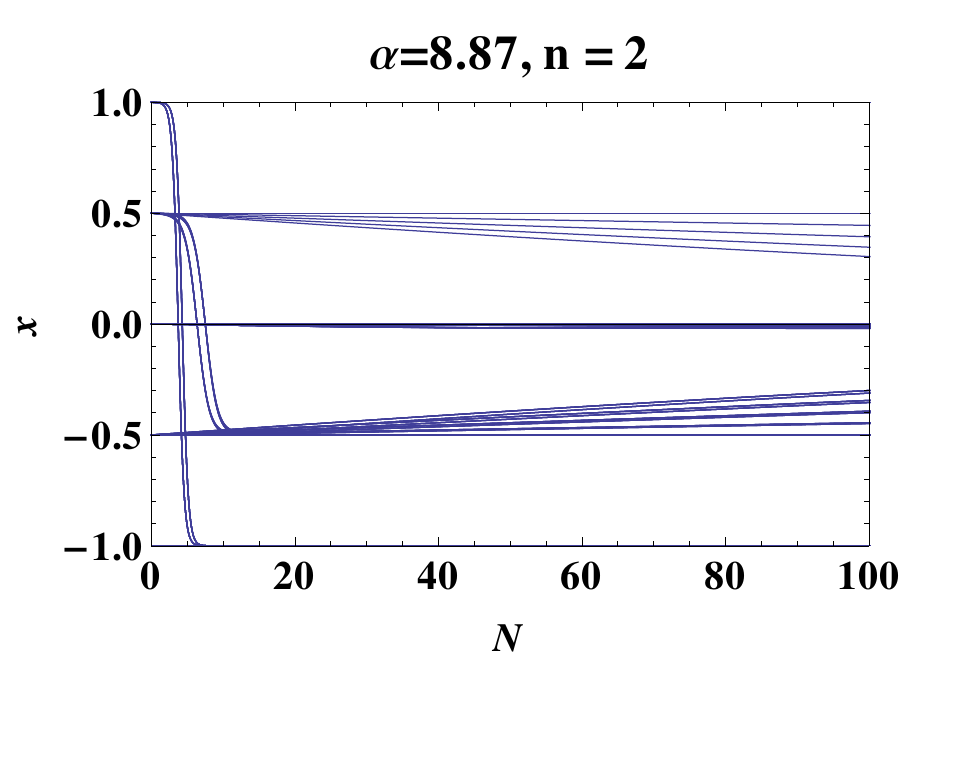}
\includegraphics[width=0.32\columnwidth]{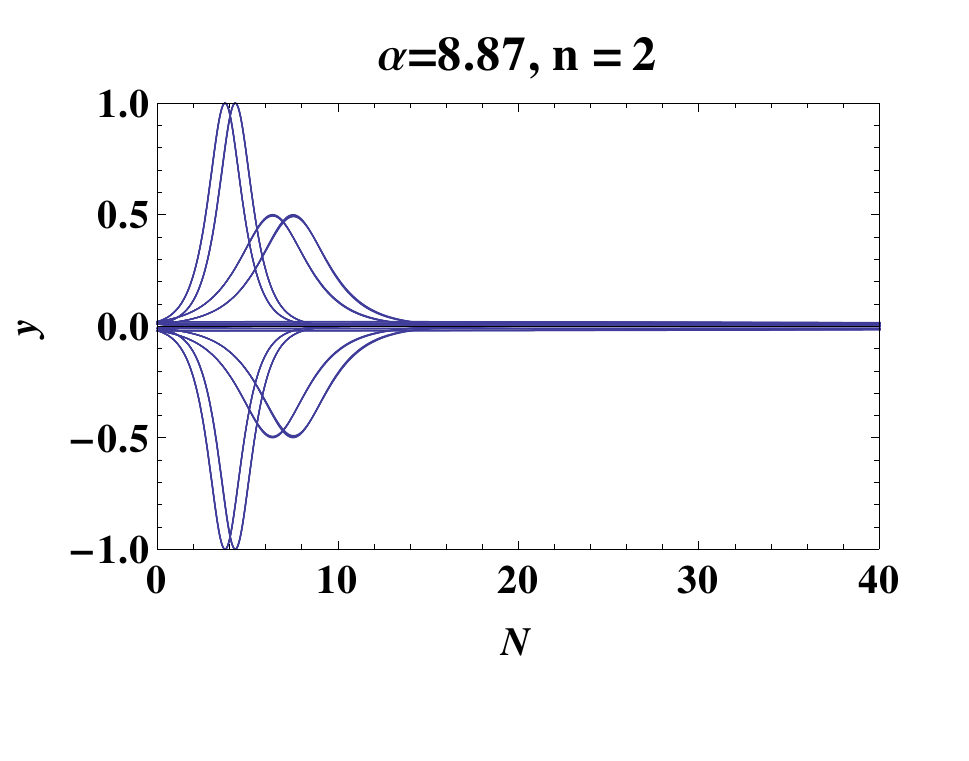}
\includegraphics[width=0.32\columnwidth]{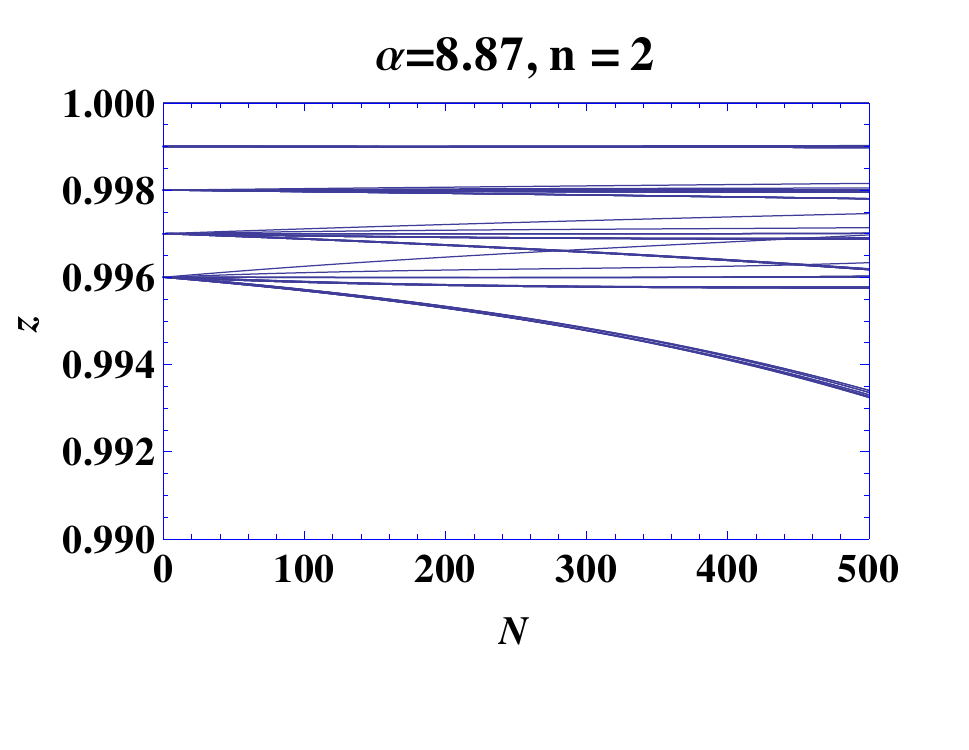}
\caption{\em Projection of perturbations along $x, y, z$ for critical point $B$ for $n=2$ and $\alpha = 8.87$}. 
\label{figb2}
\end{center}
\end{figure}
\begin{figure}[ht]
\begin{center}
\includegraphics[width=0.32\columnwidth]{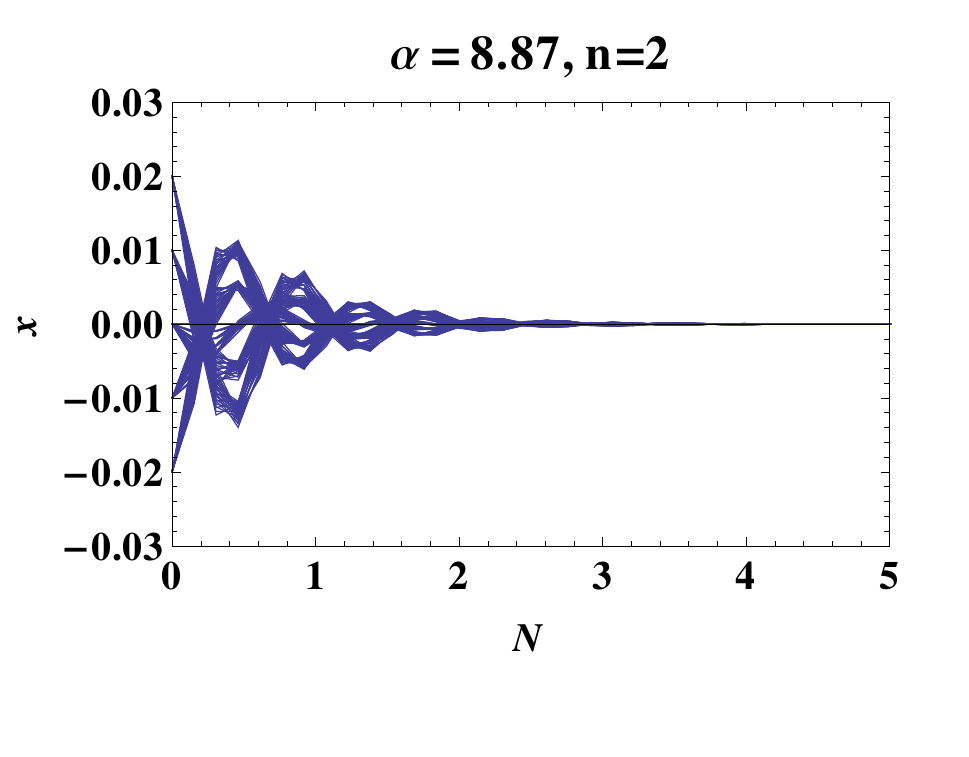}
\includegraphics[width=0.32\columnwidth]{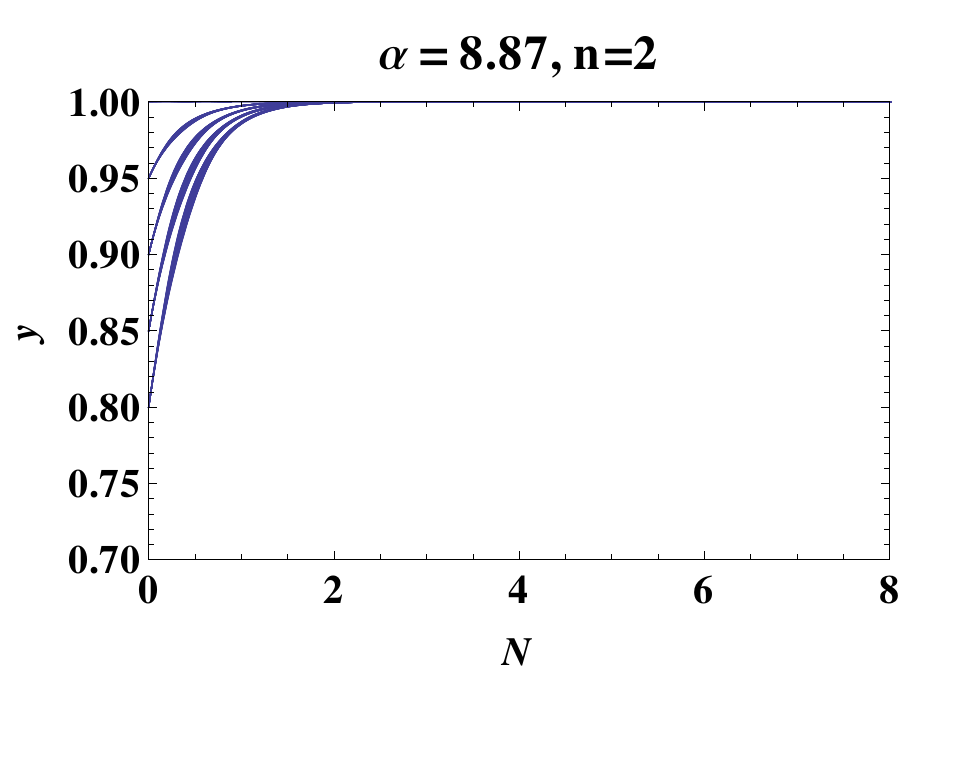}
\includegraphics[width=0.32\columnwidth]{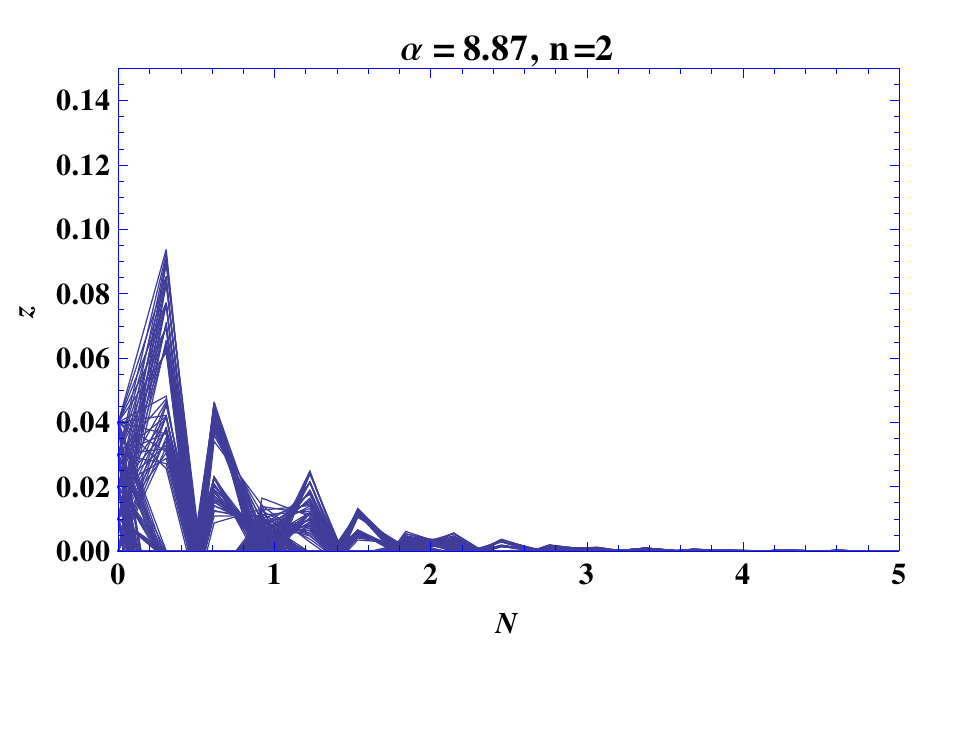}
\caption{\em Projection of perturbations along $x, y, z$ for critical point $C$ for $n=2$ and $\alpha = 8.87$.}
\label{figc2}
\end{center}
\end{figure}
\begin{figure}[ht]
\begin{center}
\includegraphics[width=0.32\columnwidth]{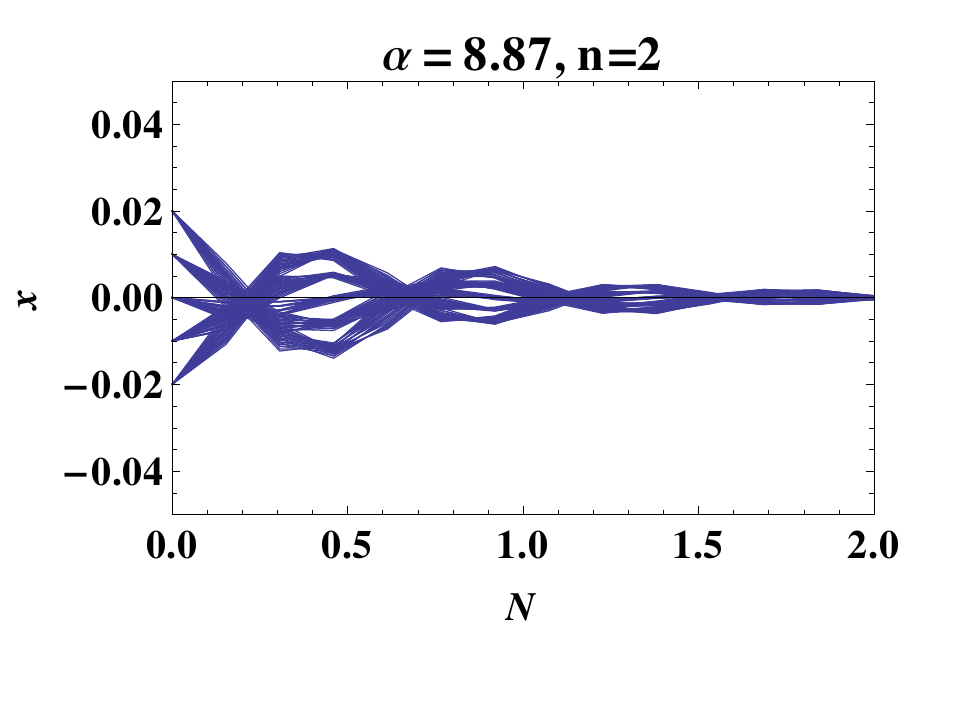}
\includegraphics[width=0.32\columnwidth]{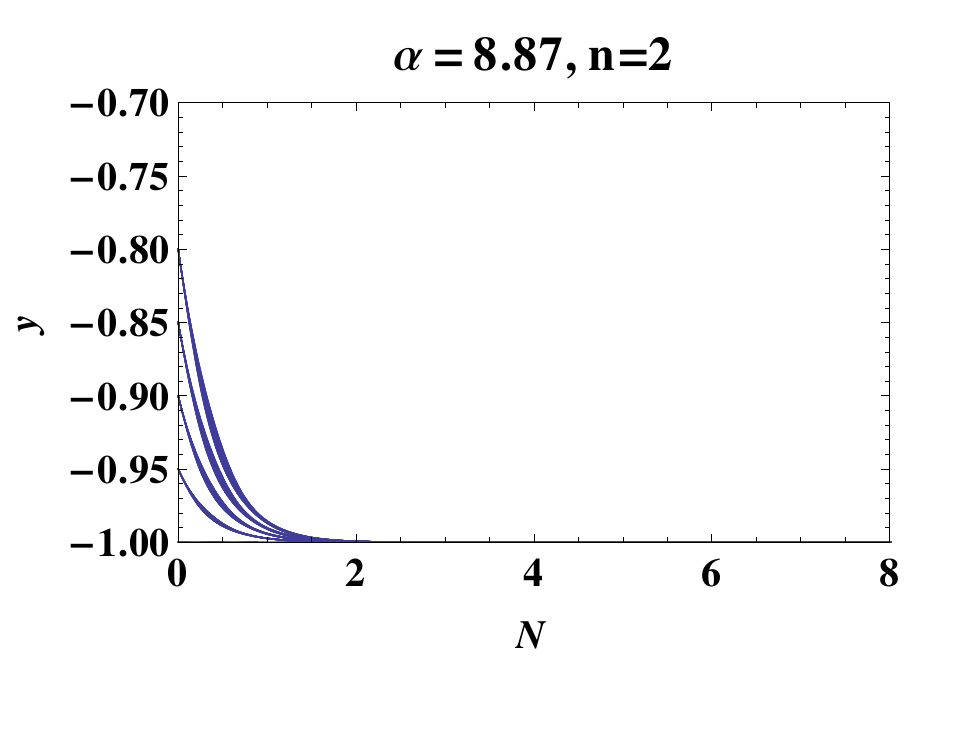}
\includegraphics[width=0.32\columnwidth]{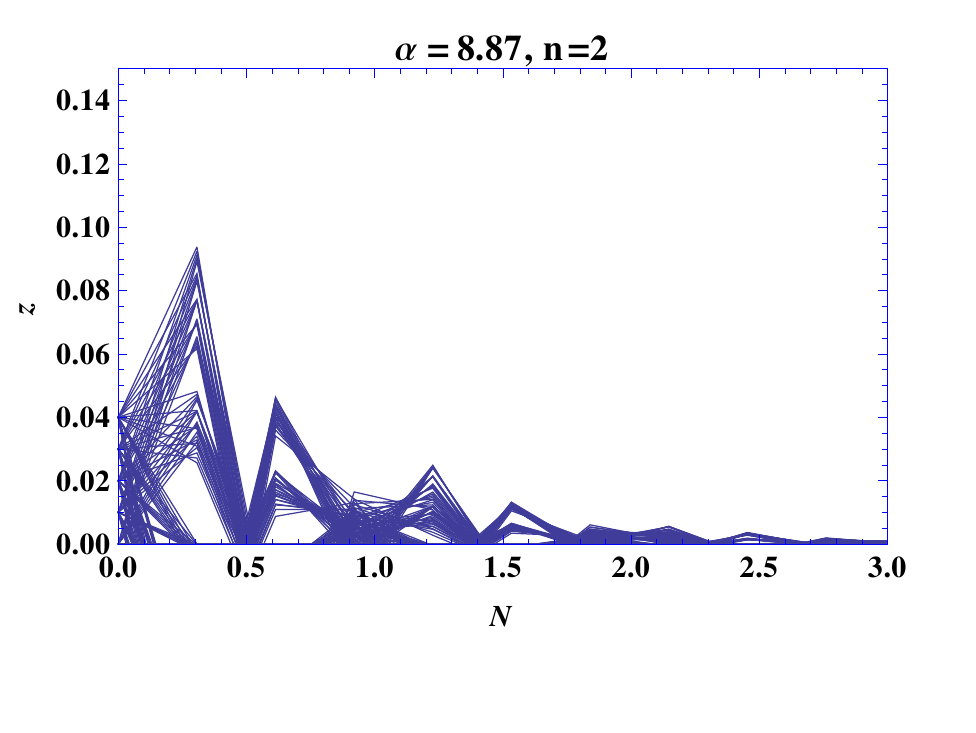}
\caption{\em Projection of perturbations along $x, y, z$ for critical point $D$ for $n=2$ and $\alpha = 8.87$.}
\label{figd2}
\end{center}
\end{figure}
Figure \ref{figb2}, fig \ref{figc2} and fig \ref{figd2} shows the numerical results near the critical points $B, C, D$ respectively for $n=2$ and $\alpha = 8.87$. One can see that for point $B$, some of the perturbations do not come back (for instance, the first and third panel of fig \ref{figb2}). This indicates that the critical point $B$ is unstable corresponding to $n=2$. However, for the points $C$ and $D$, the system is stable as seen from the figures. We have found the same conclusion using CMT in the previous section.\\ 

\par
\begin{figure}[ht]
\begin{center}
\includegraphics[width=0.32\columnwidth]{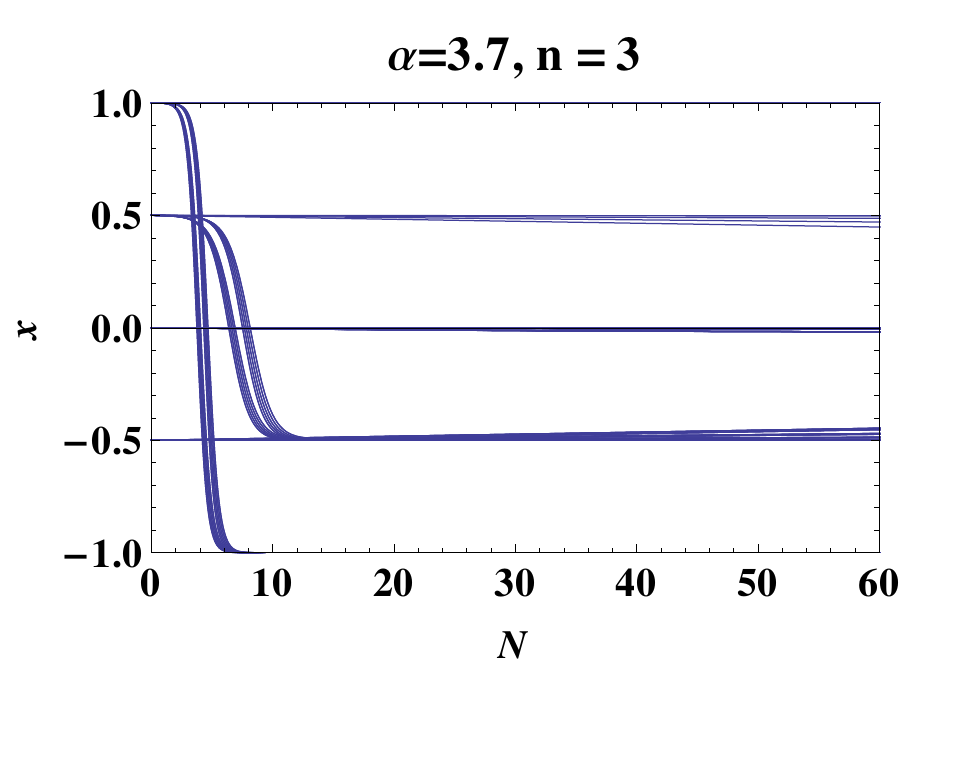}
\includegraphics[width=0.32\columnwidth]{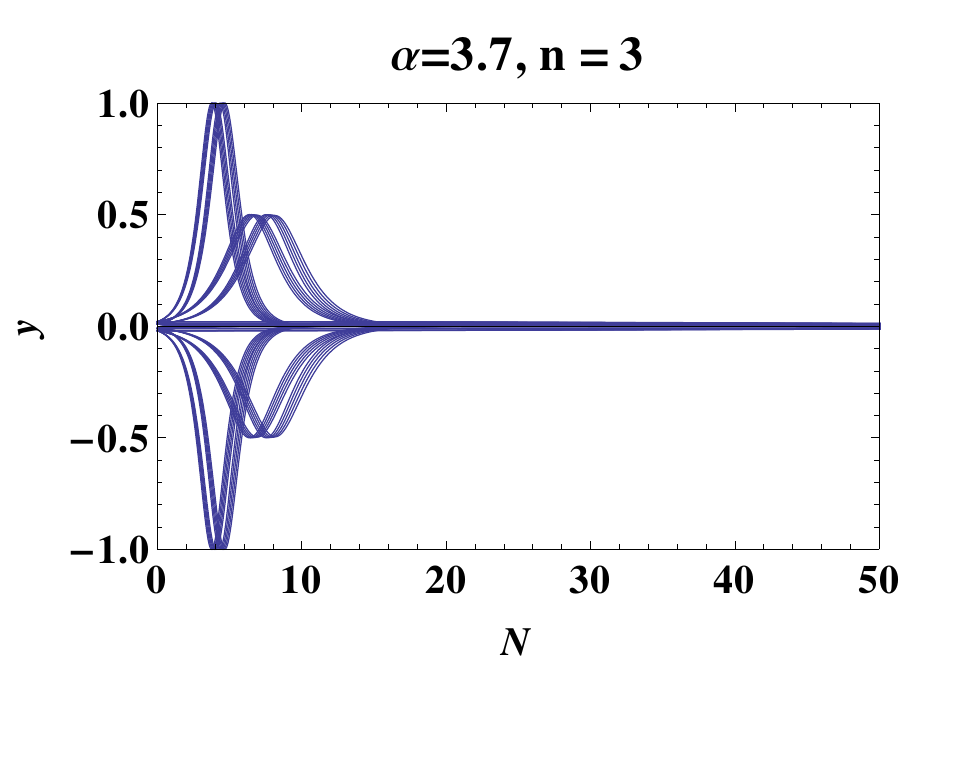}
\includegraphics[width=0.32\columnwidth]{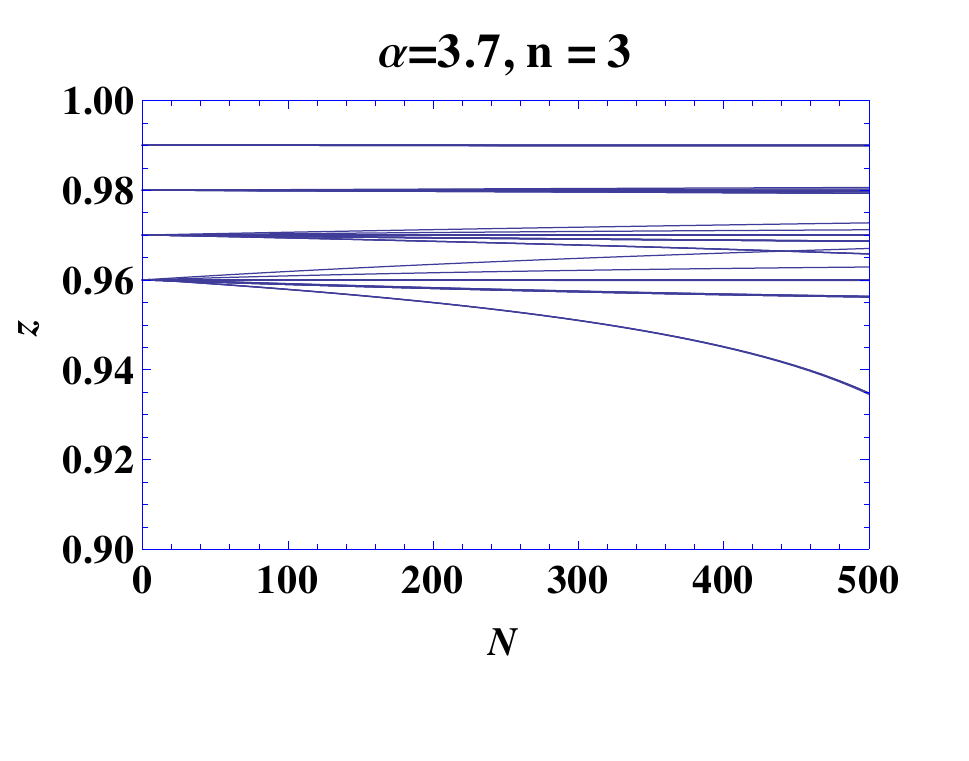}
\caption{\em Projection of perturbations along $x, y, z$ for critical point $B$ for $n=3$ and $\alpha = 3.7$.}
\label{figb3}
\end{center}
\end{figure}

\begin{figure}[ht]
\begin{center}
\includegraphics[width=0.32\columnwidth]{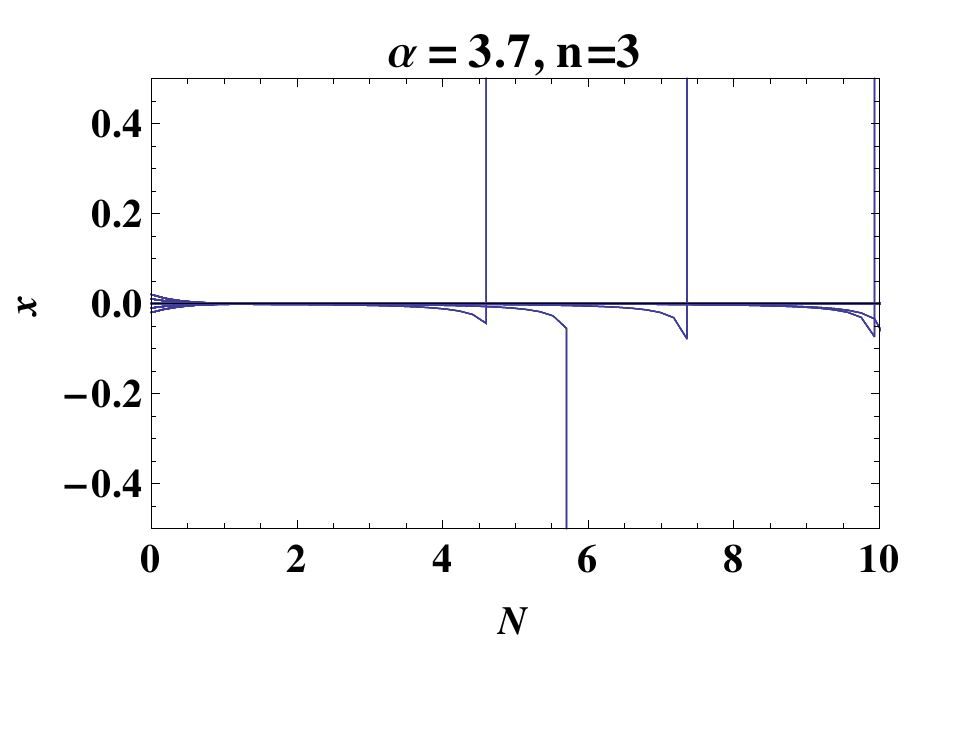}
\includegraphics[width=0.32\columnwidth]{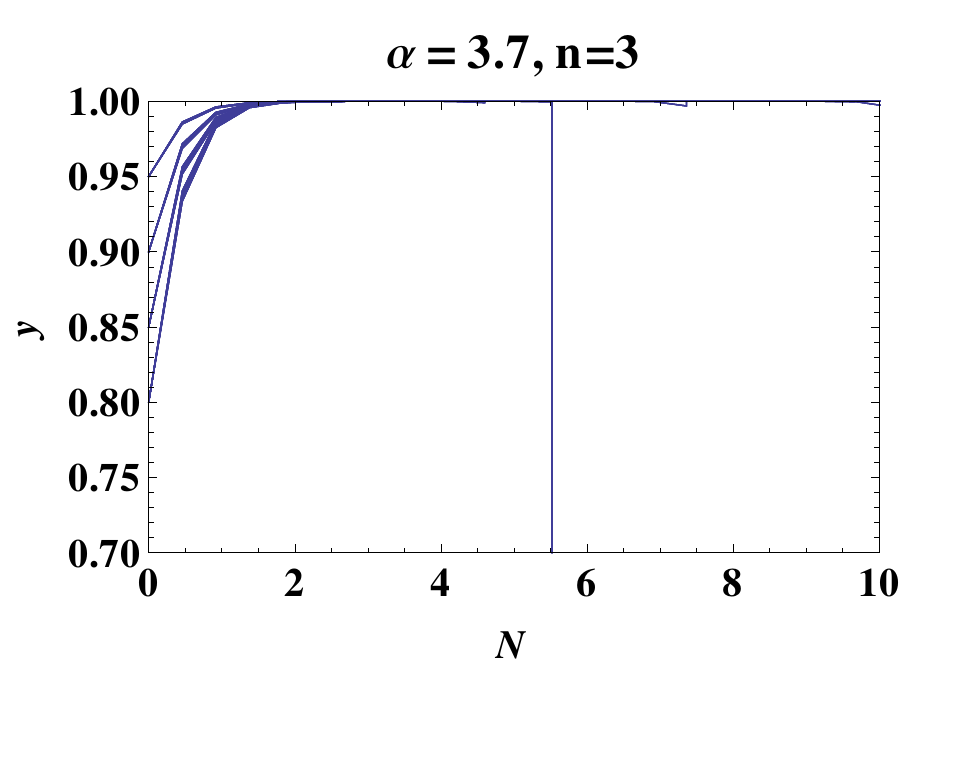}
\includegraphics[width=0.32\columnwidth]{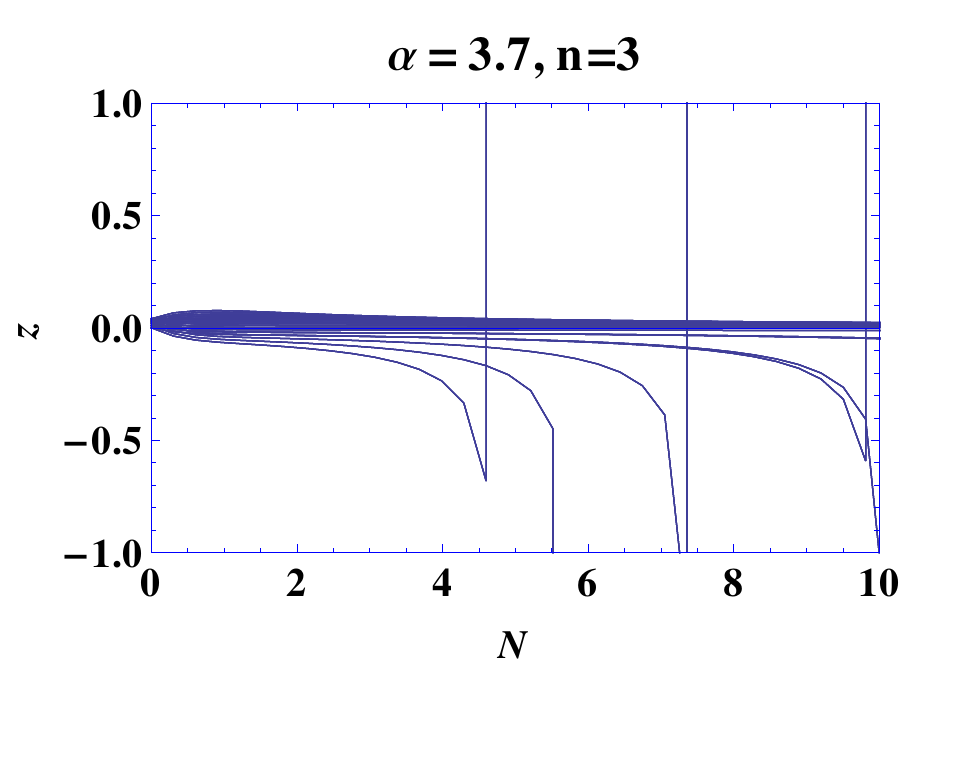}
\caption{\em Projection of perturbations along $x, y, z$ for critical point $C$ for $n=3$ and $\alpha = 3.7$.}
\label{figc3}
\end{center}
\end{figure}

\begin{figure}[ht]
\begin{center}
\includegraphics[width=0.32\columnwidth]{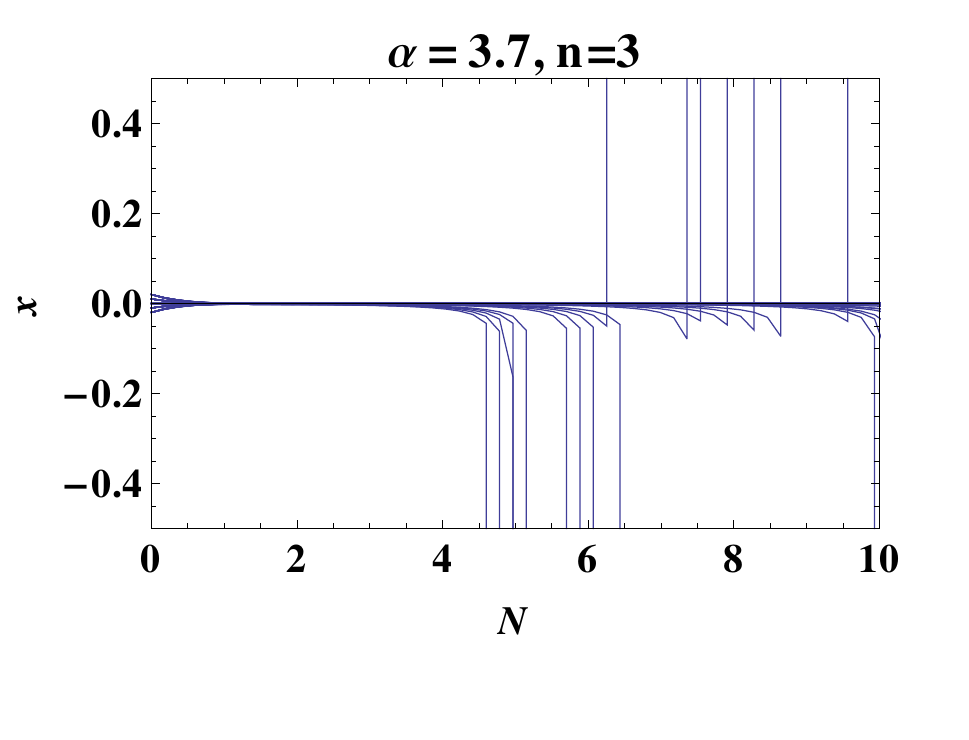}
\includegraphics[width=0.32\columnwidth]{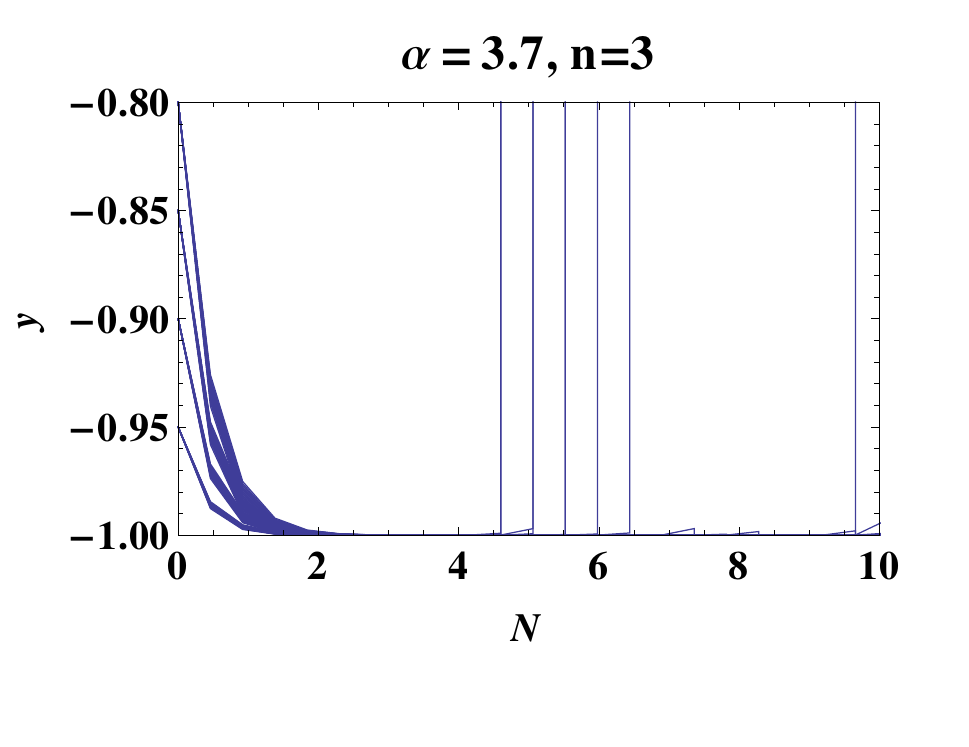}
\includegraphics[width=0.32\columnwidth]{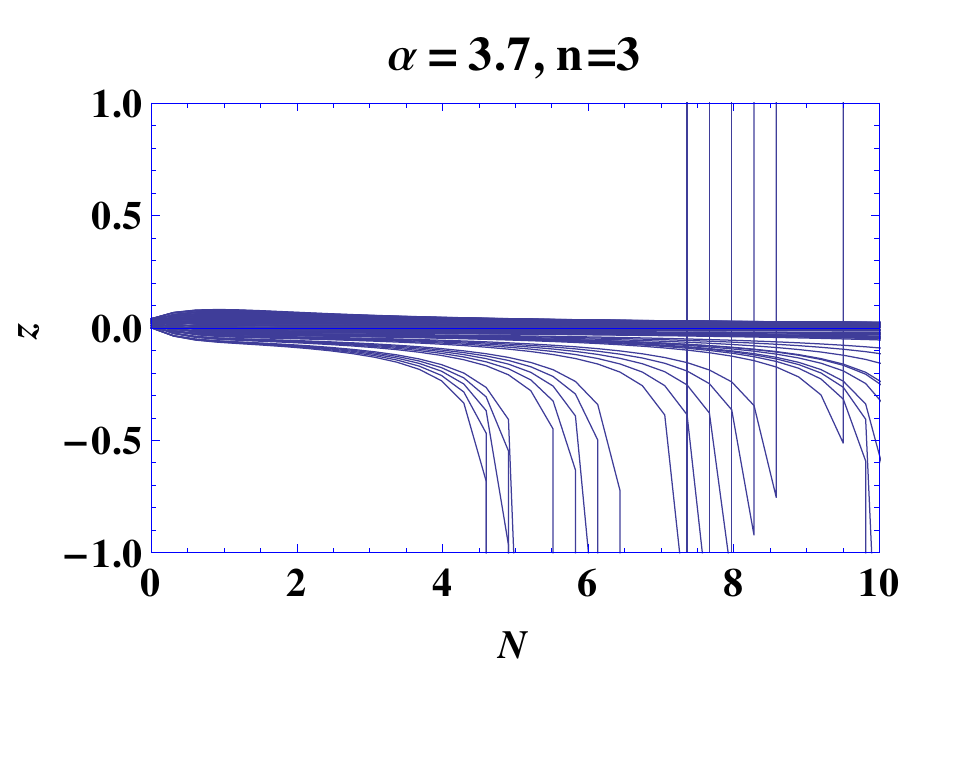}
\caption{\em Projection of perturbations along $x, y, z$ for critical point $D$ for $n=3$ and $\alpha = 3.7$.}
\label{figd3}
\end{center} 
\end{figure}

 The same analysis when applied to the critical points $B, C$ and $D$ for $n=3$ and $\alpha=3.7$, we obtain numerical results that are shown in fig \ref{figb3}, fig \ref{figc3} and fig \ref{figd3} respectively. One can see from fig \ref{figb3} that the perturbations corresponding to point $B$ exhibit the same nature as $n=2$ and thus indicates that $B$ is also an  unstable point for $n=3$.  However, for $n =3$, the critical points $C$ and $D$ show behaviour opposite to that of $n=2$, i.e, some perturbations never come back hence they are unstable in nature. This result agrees with the result obtained from CMT analysis. From CMT analysis also we have obtained that points $C$ and $D$ correspond to unstable fixed points for $n=3$.


\par
\begin{figure}[ht]
\begin{center}
\includegraphics[width=0.32\columnwidth]{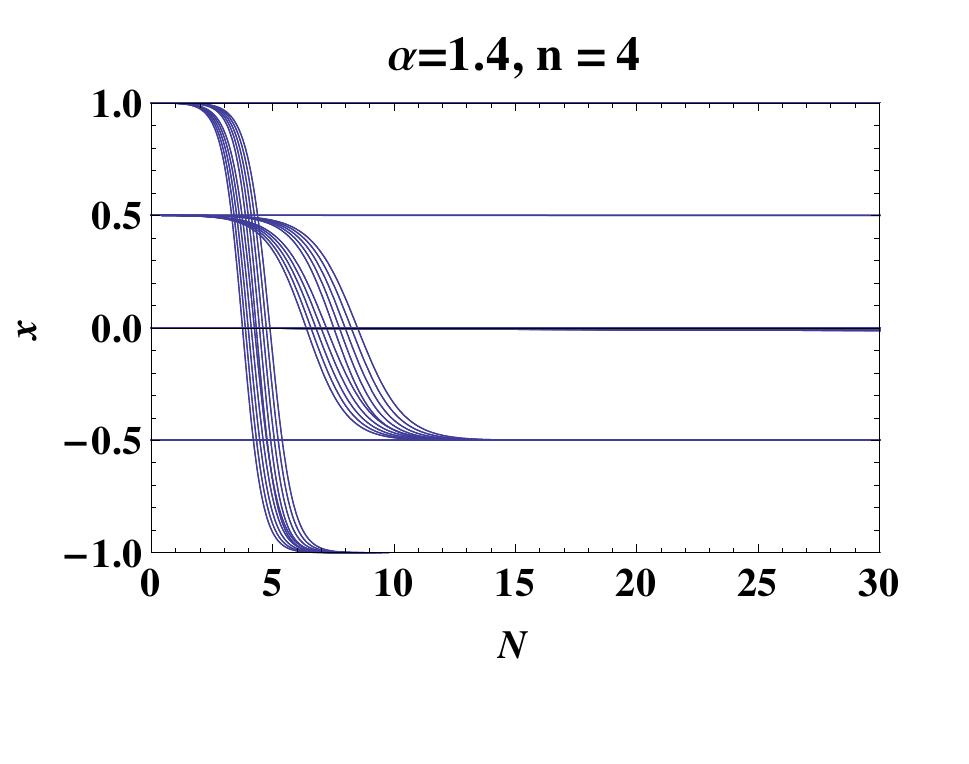}
\includegraphics[width=0.32\columnwidth]{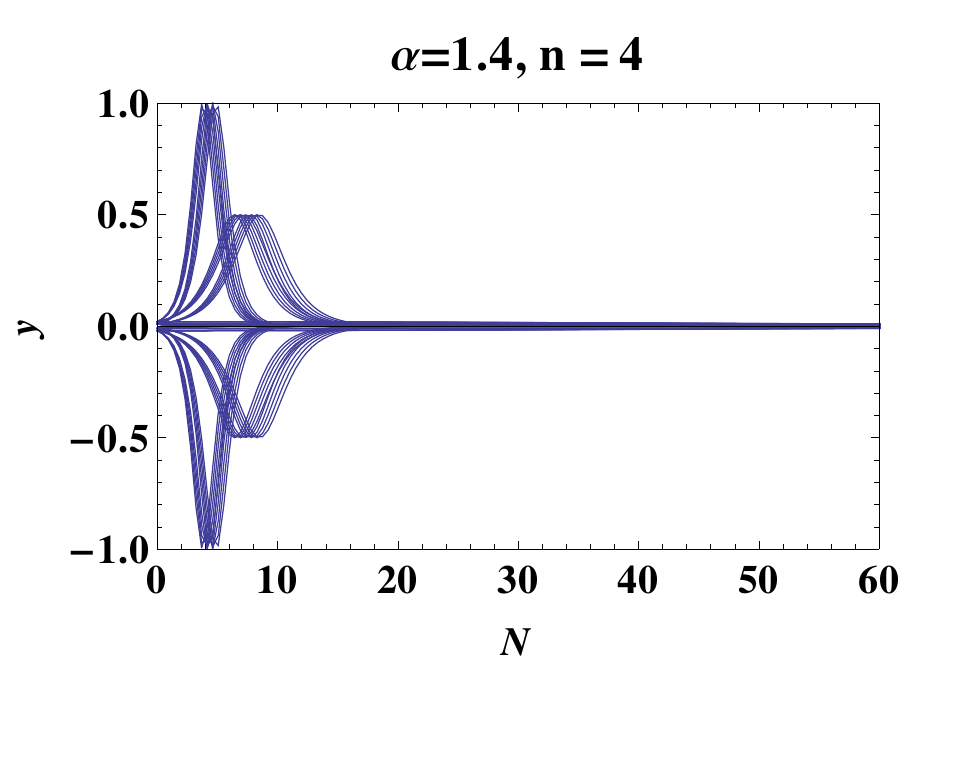}
\includegraphics[width=0.32\columnwidth]{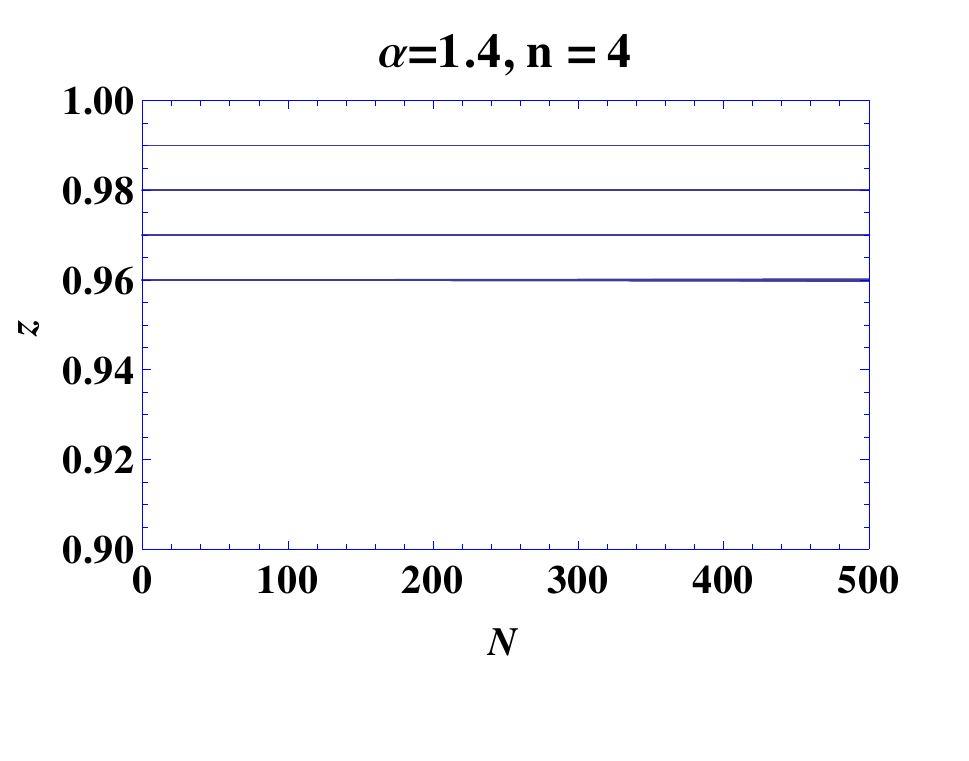}
\caption{\em Projection of perturbations along $x, y, z$ for critical point $B$ for $n=4$ and $\alpha = 1.4$.}
\label{figb4}
\end{center}
\end{figure}

\begin{figure}[ht]
\begin{center}
\includegraphics[width=0.32\columnwidth]{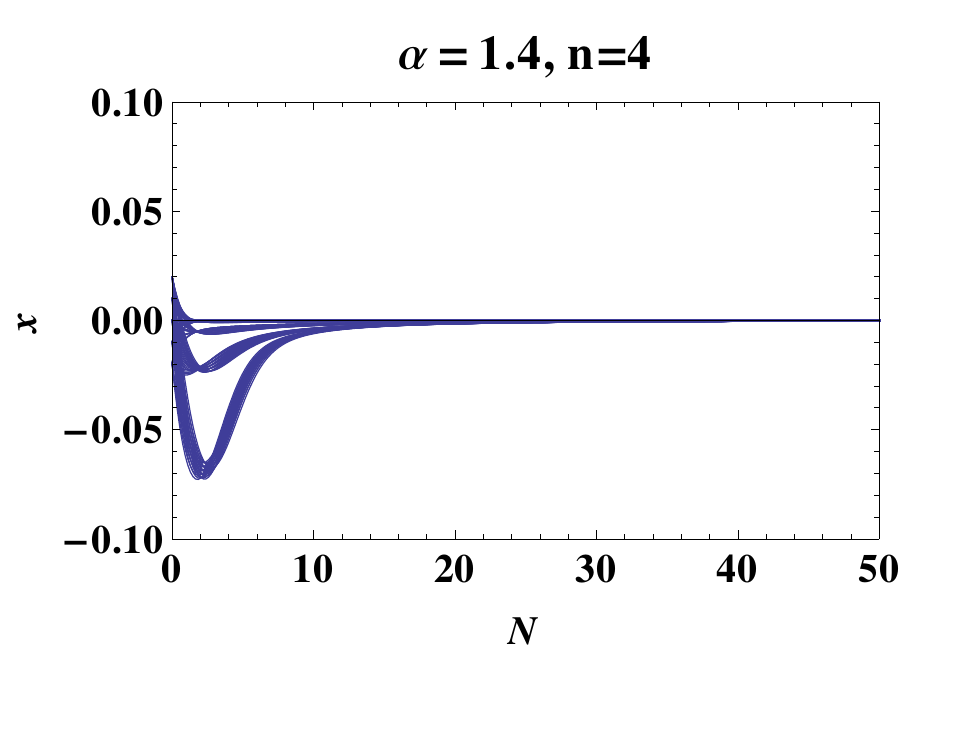}
\includegraphics[width=0.32\columnwidth]{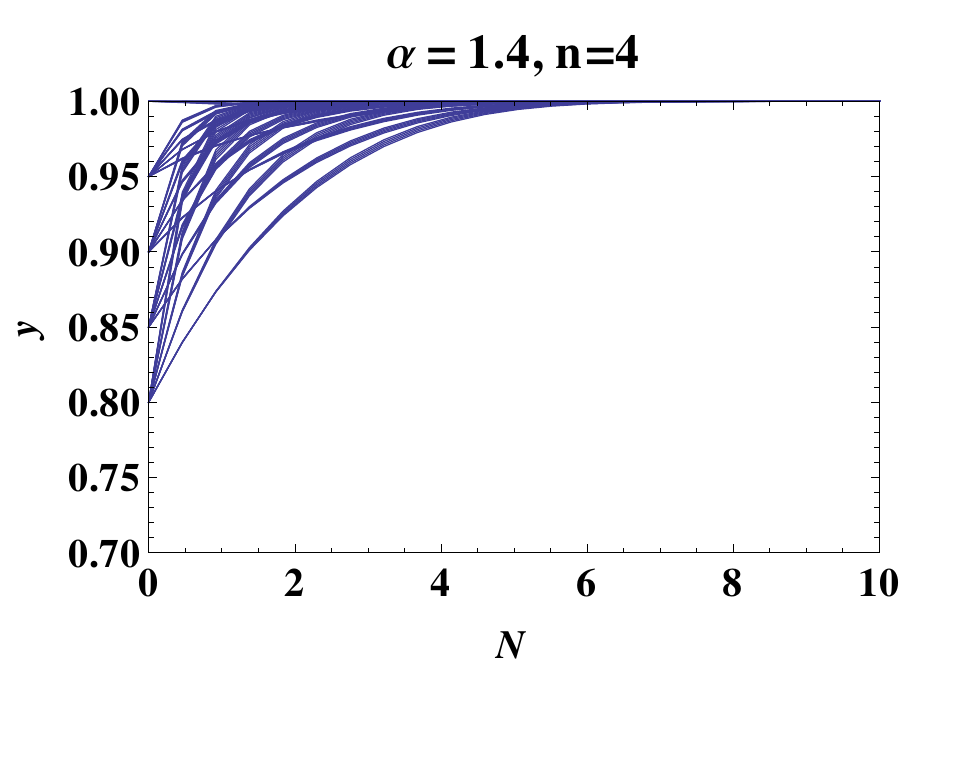}
\includegraphics[width=0.32\columnwidth]{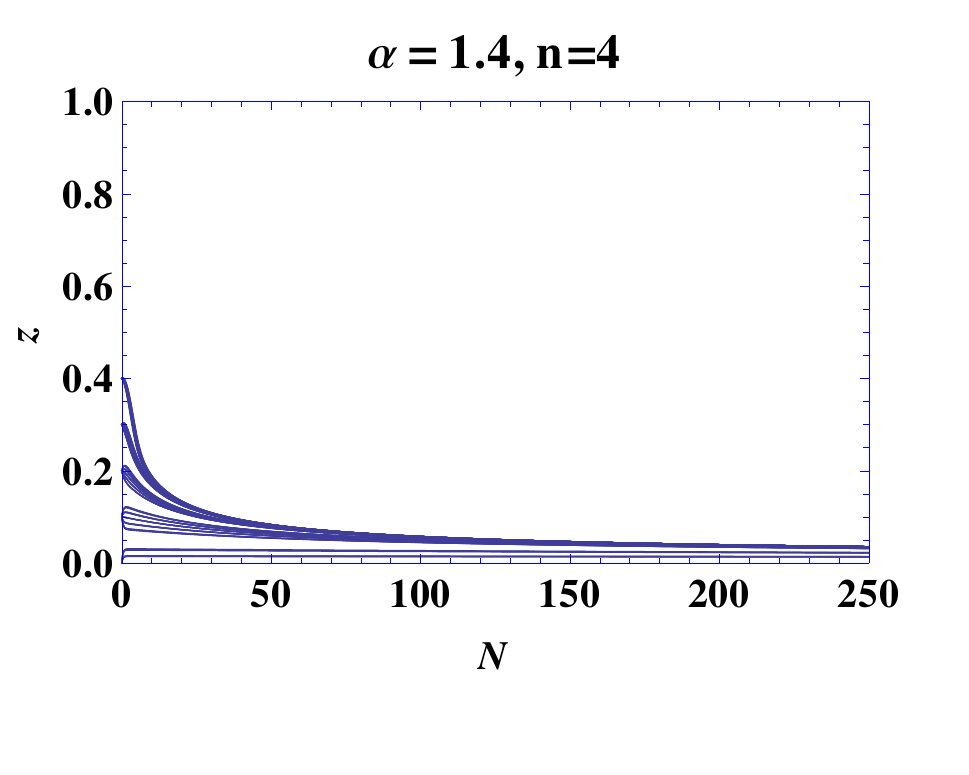}
\caption{\em Projection of perturbations along $x, y, z$ for critical point $C$ for $n=4$ and $\alpha = 1.4$.}
\label{figc4}
\end{center}
\end{figure}

\begin{figure}[ht]
\begin{center}
\includegraphics[width=0.32\columnwidth]{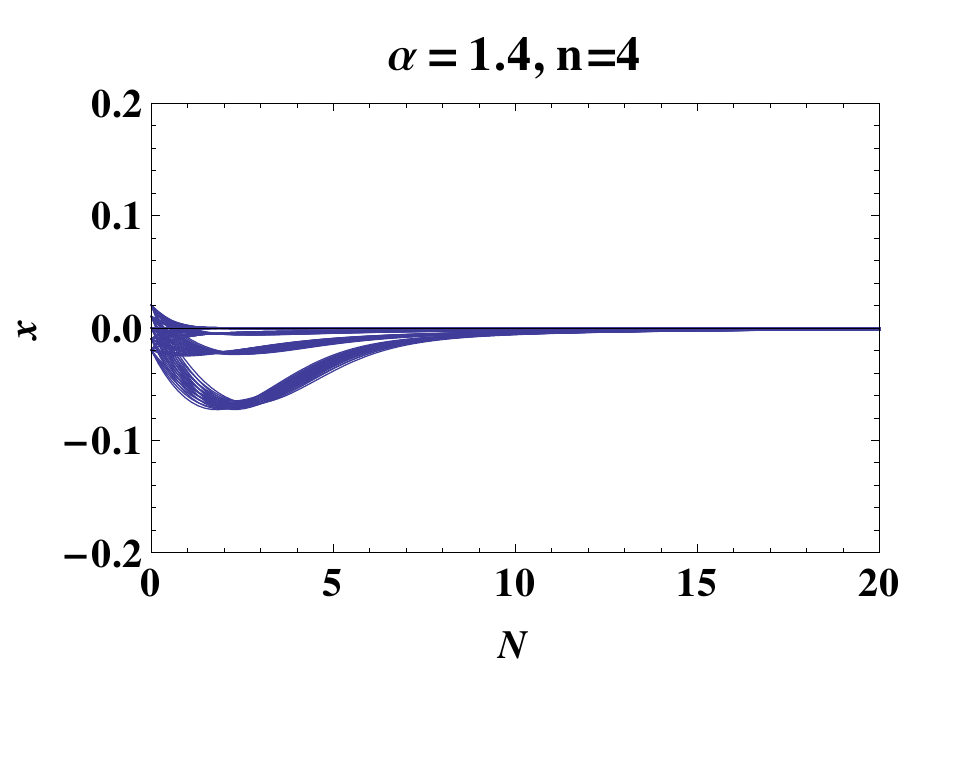}
\includegraphics[width=0.32\columnwidth]{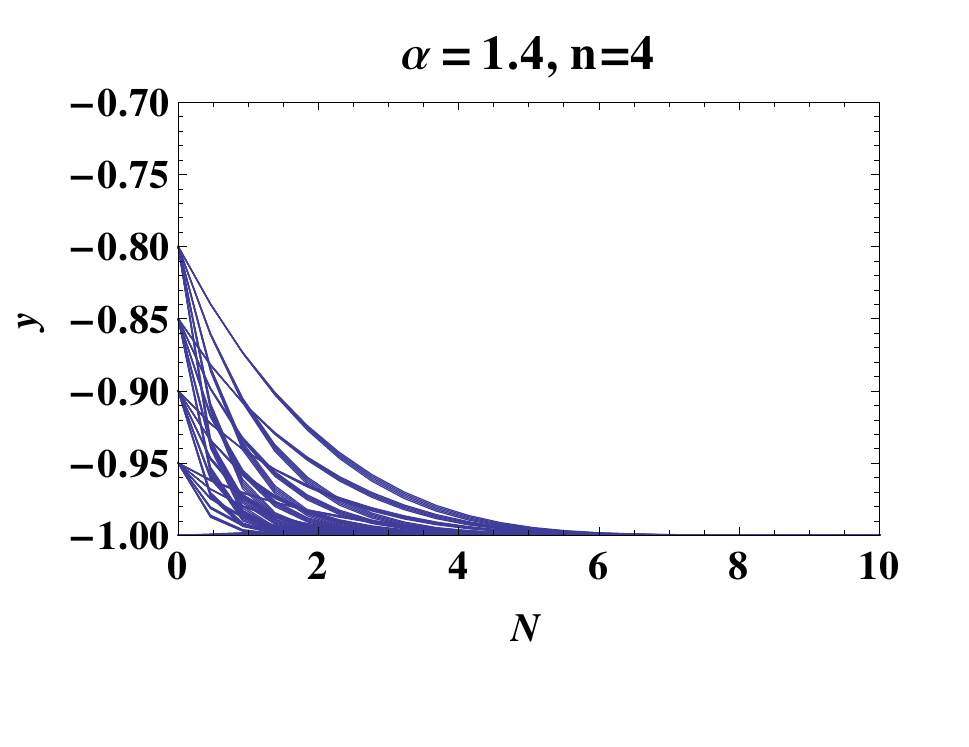}
\includegraphics[width=0.32\columnwidth]{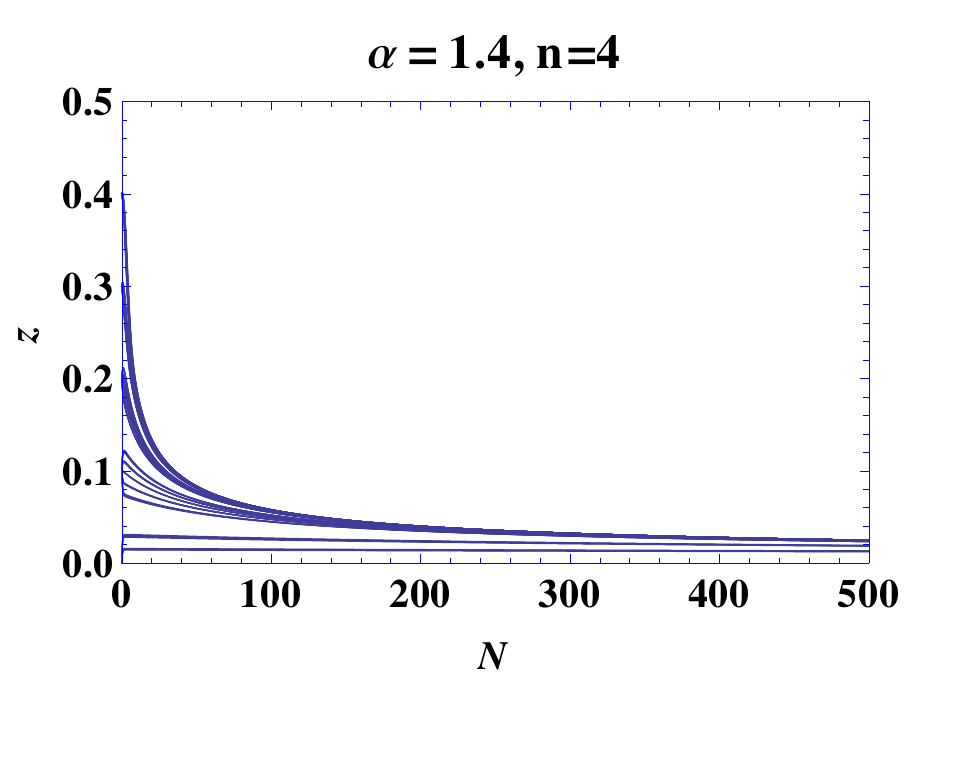}
\caption{\em Projection of perturbations along $x, y, z$ for critical point $D$ for $n=4$ and $\alpha = 1.4$.}
\label{figd4}
\end{center}
\end{figure}

 Fig \ref{figb4}, fig \ref{figc4} and fig \ref{figd4} shows the numerical results for the critical points $B, C$ and $D$ respectively for $n=4$ and $\alpha = 1.4$. Again point $B$ corresponds to an unstable fixed point whereas points $C$ and $D$ correspond to stable fixed points as evident from the figures. From CMT analysis as well, we have obtained that for $n=4$, point B is unstable and points $C$ and $D$ correspond to stable fixed points. Both the analysis implies that $n=4$ can provide a stable solution corresponding to critical point $C,D$ and may characterize the present dark energy dominated phase of the universe. 

\section{Cosmological implications}
From Table \ref{criticalpoint} it is evident that fixed point $A$ is always a saddle point. This also corresponds to a matter dominated and decelerated universe (as given by equation \ref{wphi}). According to the stability analysis, as this point is saddle in nature, it is possible that this fixed point can indeed correspond to an early phase of matter dominated era when the DE component was subdominant.

 For fixed point $B$, both from CMT analysis and figures \ref{figb2}, \ref{figb3} and \ref{figb4} it is evident that point $B$ always represent an unstable fixed point irrespective of the value of steepness index $n$. Also as $w_{\phi} = 1$ and $\Omega_m$ and $\Omega_{\phi}$ are both non-zero, this fixed point represents a state of the universe after the radiation domination, which could have contributions from both dark matter and dark energy. Though a significant contribution from the dark energy after radiation domination is not supported by observations, one can adjust the contribution of the dark energy by choosing an appropriate initial value of $x$. \\

\par Both the fixed points $C$ and $D$ eventually represent a purely dark energy dominated phase, effectively a cosmological constant ($\Omega_m = 0$, $\Omega_{\phi} = 1$ and $w_{\phi}=-1$). For $n=2$, both represent a stable solution. For $n=3$ these points are unstable fixed points. For $n=4$, these fixed points are stable in nature. From cosmological point of view these fixed points are completely dark energy dominated and accelerated. As we have considered a universe after the radiation dominated era and dominated by dark matter and dark energy, the unstable nature of these fixed points is not physically interesting. The solutions which starts from these fixed point will describe a universe which is dark energy dominated and accelerated immediately after the radiation domination. This could have serious impact on the structure formation of the universe. On the other hand stable nature of these fixed points will describe a universe which will be completely dominated by the dark energy in the future and will accelerate for ever.

\par
\begin{figure}[ht]
\begin{center}
\includegraphics[width=0.32\columnwidth]{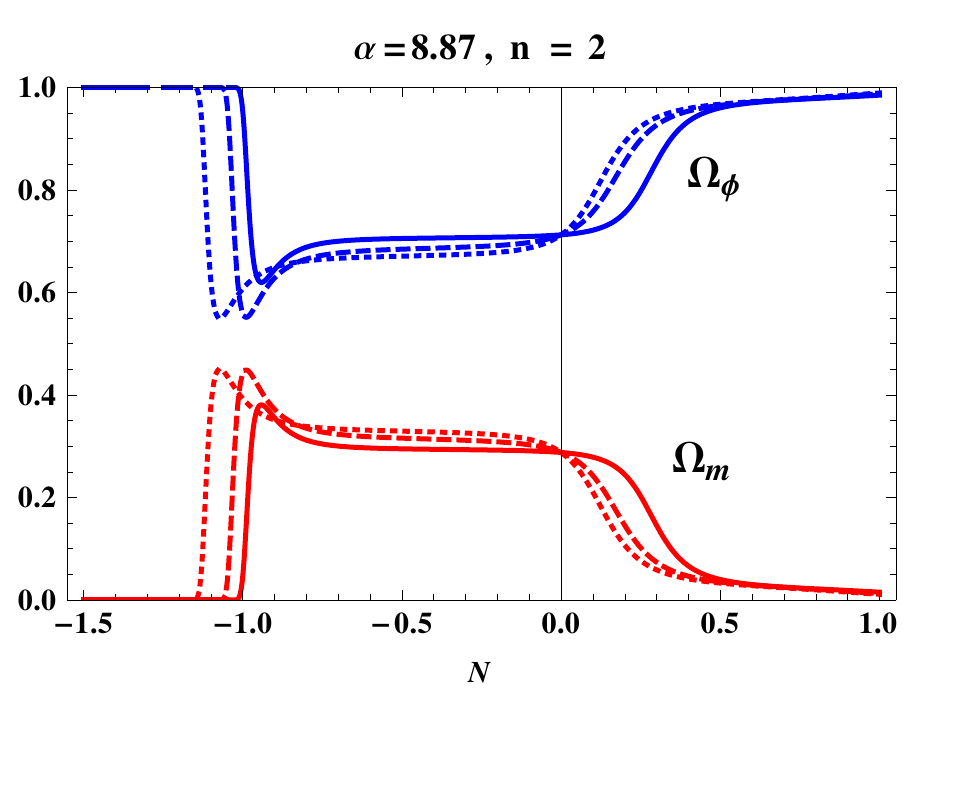}
\includegraphics[width=0.32\columnwidth]{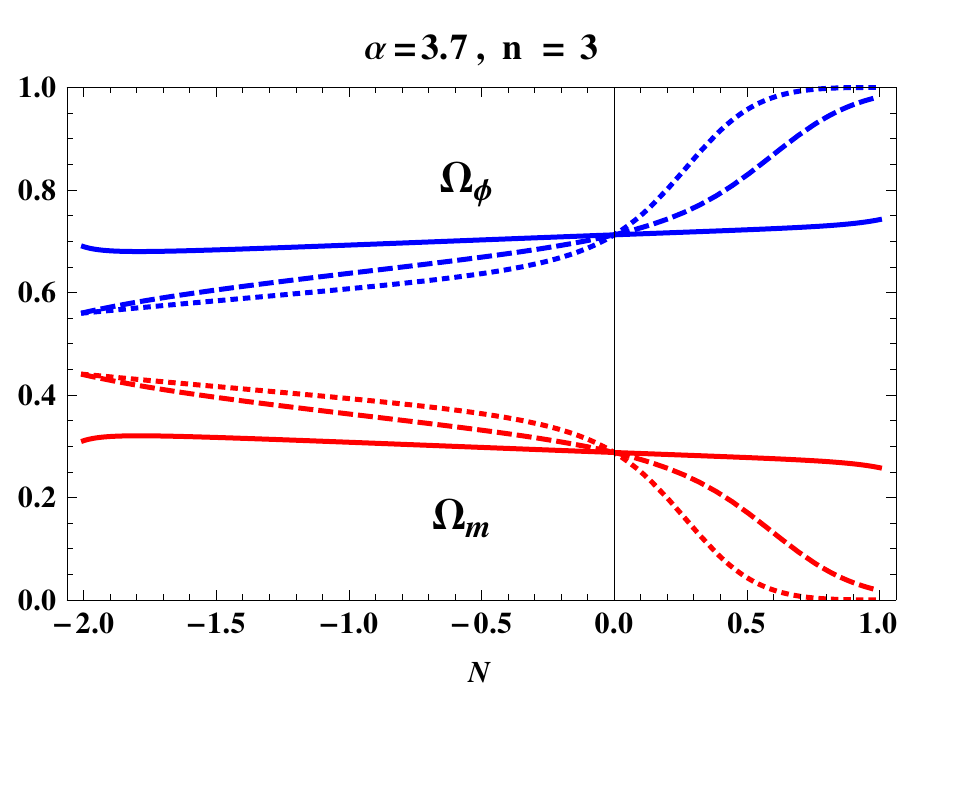}
\includegraphics[width=0.32\columnwidth]{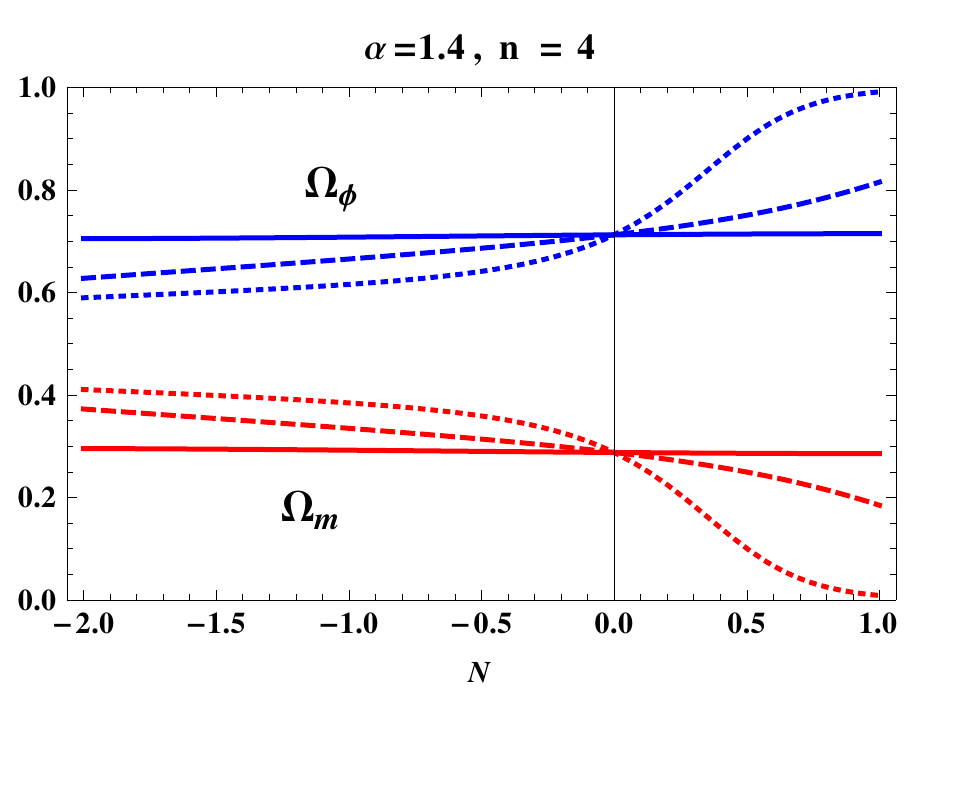}
\caption{\em Plots of $\Omega_{\phi}$ and $\Omega_m$ for different values of $z_0$ considering $x_0 = - 0.272, y_0 =  0.799 $. The blue lines represents the $\Omega_{\phi}$ and red lines are for $\Omega_m$. Solid lines correspond to $z_0 \simeq 0.8$, dashed lines correspond to  $z_0 \simeq 0.5$ and the dotted lines correspond to $z_0 \simeq 0.2$}
\label{omega}
\end{center}
\end{figure}

 In what follows we are going to check numerically if the steep potential can reproduce the evolution of the universe accordingly. To do so we have estimated the present values of $x, y$ from the present values of $\Omega_\phi$ and $w_{\phi}$. We have considered the values of $\Omega_{\phi 0} \simeq 0.714$ and $ w_{\phi 0} \simeq -0.8 $ from \cite{wali} where constraints on the cosmological parameters are found out using this particular steep potential. One can estimate the initial conditions of $x$ and $y$ from these two equations, $ \Omega_{\phi 0} = x_0 ^2 + y_0 ^2 $  and $w_{\phi 0} =  x_0 ^2 - y_0 ^2 $. Considering the above mentioned values of $w_{\phi 0}$ and $\Omega_{\phi 0}$, we obtain $x_0 = \pm 0.272, y_0 = \pm 0.799 $. To solve the equations we need to give also the present value of $z$. By construction, $0 \leq z \leq 1$; so we have considered three different values of $z$ as the present value. As mentioned earlier we have considered values of $\alpha$ from the best fit values of \cite{wali}
i.e, for $n=2, \alpha = 8.87$, for $n=3, \alpha = 3.7$ and for $n=4, \alpha = 1.4$. The plots of $\Omega_{\phi}$ and $\Omega_m$ are given in the Fig \ref{omega}. One can see from Fig \ref{omega} that for almost any value of $z_0$ the universe is always dominated by the dark energy and there is no dark matter-dark energy cross over.  Though consideration of this potential can explain the present accelerated expansion of the universe it may have serious impact on the throughout evolution of the universe.

\section{Conclusion}

In this work we have performed stability analysis for steep potential $V(\phi) = V_0 e^{\alpha(\frac{\phi}{M_p})^n}$ which has been recently introduced in \cite{wali} to explain both the inflation and present day accelerated expansion. Though a thorough study of cosmological dynamics of this potential already exists in the literature, the dynamical systems analysis of it can reveal more about its qualitative nature. The study has been done for three different values of $n = 2, 3, 4$ and the value of the $\alpha$ parameter has been considered from the best fit values of $\alpha$ in \cite{wali} to keep our study close to observations.

To do the dynamical  systems analysis of the system we have written the system as a set of autonomous equations using suitable variable transformations. Due to singular nature of the Jacobian matrix for some fixed points we have considered another transformation of the variable $z$ which regularizes the system so that the techniques of the dynamical systems analysis can be applied. The fixed points are found out for the new system and all of them are nonhyperbolic in nature. We have used both analytical and numerical tools to find stability of these fixed point. To study it analytically Centre Manifold Theorem (CMT) has been used. For the numerical study a perturbation technique has been employed. In this technique the system is perturbed in the neighbourhood of the desired fixed point and allowed to evolve. As the time passes if the system comes back to the fixed point then the fixed point is stable otherwise unstable. Both CMT and perturbation technique give us the same result. 

\par To summarize the results, as the beginning the fixed point $A$ is most favorable. Fixed point $B$ can also serve as the beginning for a choice of initial value of $x$, $x_0 \ll 1$ until it does not have any conflict with the observations. For $n=2$ there are two possible ultimate fates of the universe, fixed points $C$ and $D$. For $n=3$ there may or may not be any late time attractor. For $n=4$ the universe will approach either $C$ or $D$. So in this model the universe has qualitatively only one future which is the dark energy domination exhibiting accelerated expansion. We have also studied the throughout evolution of the universe considering this potential. One interesting finding is that the consideration of this potential makes the universe dark energy dominated in the very remote past and there is no dark matter and dark energy domination flip. So though this potential can explain the present accelerated expansion very well, considering the throughout evolution of it, these steep exponential potentials do not seem to be a good candidate for dark energy.

\section{Acknowledgement}
SD and MB acknowledge the financial support from SERB, DST, Government of India through the project EMR/2016/007162. SD would also like to acknowledge IUCAA, Pune for providing support through associateship programme. Authors are also thankful to Prof. Pijush Kanti Ghosh, Prof. Madan Mohan Panja and Dr. Sreekantha Sil for useful discussions. We are thankful to the anonymous referee for the useful suggestions which have improved the quality of the paper.
\newpage
\newpage
\begin{center}
    \bf{APPENDIX}
    \newline{\bf\large{Detailed analysis using Centre Manifold Theory}}
\end{center}
\appendix
\section{For point B}\label{appendixA1}
As Point B is a non-hyperbolic critical point having zero eigen value, one can not arrive at any conclusion about the stability of the point using linear stability theory. The critical point B is (Any, $0, 1$). For ease of calculation, let us consider the point as ($\pm a$, 0, 1) where $ 0 \le a \le 1 $. \\
To proceed we have to first transform the system of equations (\ref{xprimezfinal})-(\ref{zprimezfinal}) into the standard form 
as mentioned in section \ref{CMT}. Firstly, for this point, the coordinates are rescaled such that
\begin{equation}
X=x\mp a
\end{equation}
\begin{equation}
Y=y
\end{equation}
\begin{equation}
Z=z-1
\end{equation}
This rescaling moves the point (${\pm a},0, 1$) to the origin ($0, 0, 0$) of the
phase space.
 
Then Eqs. (\ref{xprimezfinal})-(\ref{zprimezfinal}) becomes
\begin{equation}
    X'=\frac{1}{2}(-Z)^{n-1} \{3(X\pm a)^3-3(X\pm a)(Y^2+1)\}-\sqrt{\frac{3}{2}}Y^2(Z+1)^{n-1}
\end{equation}
\begin{equation}
    Y'=-\frac{1}{2}Y(-Z)^{n-1} \{-3(X\pm a)^2+3(Y^2-1)\}+\sqrt{\frac{3}{2}}(X \pm a)Y(Z+1)^{n-1}
\end{equation}
\begin{equation}
    Z'=\sqrt{6}(X \pm a)(n\alpha)^\frac{1}{n-1} (-Z)^{n+1}
\end{equation}
The Jacobian matrix of this system of equations evaluated at $(0,0,0)$ is obtained as  
 \begin{equation}
   J_1=    \begin{pmatrix}
      0~~ & 0~~ & -\frac{3}{2}\lbrace(\pm a)^3-(\pm a)\rbrace\\
       0~~ & \pm \sqrt{\frac{3}{2}}a~~ &0 \\
       0~~ & 0~~ & 0
    \end{pmatrix}      ~~~~~\mathrm{for}~~  n=2
  \end{equation}
 
\begin{equation}
   J_2=    \begin{pmatrix}
      0~~ & 0~~ & 0\\
       0~~ & \pm \sqrt{\frac{3}{2}}a~~ &0 \\
       0~~ & 0~~ & 0
    \end{pmatrix}      ~~~~~\mathrm{for}~~  n=3,4,5...
  \end{equation}
 One can see that for $ a=0,1$, the Jacobians $J_1$ and $J_2$ are identical. Thus we can write 
 \begin{equation}
   J_3=    \begin{pmatrix}
      0~~ & 0~~ & -\frac{3}{2}\lbrace(\pm a)^3-(\pm a)\rbrace\\
       0~~ & \pm \sqrt{\frac{3}{2}}a~~ &0 \\
       0~~ & 0~~ & 0
    \end{pmatrix}      ~~~~~\mathrm{for}~~  n=2, ~~a \ne 0,1
  \end{equation}
 
\begin{equation}
   J_4=    \begin{pmatrix}
      0~~ & 0~~ & 0\\
       0~~ & \pm \sqrt{\frac{3}{2}}a~~ &0 \\
       0~~ & 0~~ & 0
    \end{pmatrix}      ~~~~~\mathrm{for}~~ \begin{cases} n=3,4,5..., ~~a = \mathrm{any} \\
                                            n=2, ~~a = 0,1
                                            \end{cases}
  \end{equation}
 \subsection {For $n=2, ~a=0$ }\label{appendixA02} 
 In this case, the Jacobian $J_4$ becomes a null matrix and thus one cannot apply CMT theory to analyse the nature of critical points. So for $n=2, ~a=0$, CMT theory fails.\\
 
 \subsection {For $n=2, ~a=1$ }\label{appendixA2} 
 As the system of equations are not in the standard form, we make a change of variables as 
 $$X=u$$
 $$Y=w$$
 $$Z=v$$
 With this substitution above set of equations$(A.4) - (A.6)$ take the final form as
 \begin{equation}
     u'=0 + g_1(u,v,w)
 \end{equation}
 \begin{equation}
     v'=0 + g_2(u,v,w)
 \end{equation}
 \begin{equation}
     w'=\pm\sqrt{\frac{3}{2}}w + f(u,v,w)
 \end{equation}
 with the polynomials $g_1$, $g_2$ and $f$ given by
  $$ g_1=-\frac{3}{2}v(u^3\pm 3u^2+2u-uw^2-u \mp w^2)-\sqrt{\frac{3}{2}}w^2(v+1)$$
  $$g_2=-2\sqrt{6}\alpha v^3(u \pm 1)$$
  $$f=\frac{3}{2}wv[(w^2-1)-(u\pm 1)^2]+\sqrt{\frac{3}{2}}wv(u \pm 1)+\sqrt{\frac{3}{2}}wu$$
In terms of these new set of variables, the system of equations can also be represented in matrix form as
   \begin{equation}
  \begin{pmatrix}

  u'\\
  v'\\
  w'\\
    \end{pmatrix}  
    =
    \begin{pmatrix}
   0 & 0& 0\\
    
    0 &0&0\\
    0& 0& \pm \sqrt{\frac{3}{2}}&\\

    \end{pmatrix}
    \begin{pmatrix}
  u\\
  v\\
  w\\
    \end{pmatrix}  
   +
\begin{pmatrix}
  g_1\\
  g_2\\
  f\\
    \end{pmatrix} 
    \end{equation}
Now the coordinates which correspond to non-zero eigenvalues $(u, v)$ can be approximated in terms of $w$ by the functions $h_1(w)$ and $h_2(w)$ respectively as 

\begin{equation}\label{h1b}
    h_1(w)=a_2w^2+a_3w^3+O(w^4)
\end{equation}
    \begin{equation}\label{h2b}
    h_2(w)=b_2w^2+b_3w^3+O(w^4)
\end{equation}

The quasilinear partial differential equation, which the vector of functions
\begin{equation*}
    h=\begin{pmatrix}
      h_1\\
      h_2
    \end{pmatrix}
\end{equation*}

has to satisfy, is given by
\begin{equation}\label{adhw}
    Dh(w)[Aw+F(w,h(w))]-Bh(w)-g(w,h(w))=0
\end{equation}
where
\begin{equation}
g=\begin{pmatrix}
  g_1\\
  g_2
\end{pmatrix}
  ,  B = \begin{pmatrix}
      0 & 0\\
       0 & 0
    \end{pmatrix}
    ,~~F = f,~~A = \pm \sqrt{\frac{3}{2}}
\end{equation}
Next we solve equation (\ref{adhw}) by substituting $A, h, F, B, g$ into it and equating equal powers of $w$ in order to obtain $h(w)$ up to the desired order. By comparing powers of $w$ from both sides of equation (\ref{adhw}), we obtain the constants $a_2$, $a_3$, $b_2$ and $b_3$ as
    \begin{equation}
    a_2= \mp \frac{1}{2},~~ a_3=0,~~b_2=0,~~b_3=0
\end{equation}
Finally the dynamics of the system restricted to the centre manifold is given by
\begin{equation}\label{adotw}
    \Dot{w}=Aw + F(w, h(w)) = \pm \sqrt{\frac{3}{2}}w-\frac{1}{2}\sqrt{\frac{3}{2}}w^3+O(w^4)
\end{equation}
At the lowest order of equation (\ref{adotw}) we have obtained an odd-parity term with coefficient $\pm \sqrt{\frac{3}{2}}$. According to CMT theory, if the lowest order is odd-parity term with positive co-efficient then the system is unstable and if the co-efficient of the lowest order odd-parity term is negative, then the system is stable. \\
Thus for $n=2$ point  $B$ will be stable for negative values of $x$ and will be unstable for positive values of $x$.  

\subsection {For $n=3$, $a = any$ }\label{appendixA3}

 For $n=3$ also, the system of equations are not in the required standard form and thus we make a change of variable as 
 $$X=u$$
 $$Y=w$$
 $$Z=v$$
 With this substitution above set of equations $(A.4) - (A.6)$ take the final form as
 \begin{equation}
     u'=0 + g_1(u,v,w)
 \end{equation}
 \begin{equation}
     v'=0 + g_2(u,v,w)
 \end{equation}
 \begin{equation}
     w'=\pm \sqrt{\frac{3}{2}}aw + f(u,v,w)
 \end{equation}
 with the polynomials $g_1$, $g_2$ and $f$ given by
  $$ g_1=\frac{3}{2}v^2[(u \pm a)^3-(u \pm a)(w^2+1)]-\sqrt{\frac{3}{2}}w^2(v+1)^2$$
  $$g_2=\sqrt{6}(3\alpha)^\frac{1}{2} v^4(u\pm a)$$
  $$f=\pm \sqrt{ \frac{3}{2}}aw(v^2+2v)+\sqrt{\frac{3}{2}}uw(v+1)^2-\frac{3}{2}v^2w[(w^2-1)-(u\pm a)^2)]$$
In terms of these new set of variables, the system of equations can also be represented in matrix form as
   \begin{equation}
  \begin{pmatrix}

  u'\\
  v'\\
  w'\\
    \end{pmatrix}  
    =
    \begin{pmatrix}
   0 & 0& 0\\
    
    0 &0&0\\
    0& 0& \pm \sqrt{\frac{3}{2}} a&\\

    \end{pmatrix}
    \begin{pmatrix}
  u\\
  v\\
  w\\
    \end{pmatrix}  
   +
\begin{pmatrix}
  g_1\\
  g_2\\
  f\\
    \end{pmatrix} 
    \end{equation}
Now the coordinates which correspond to non-zero eigenvalues $(u, v)$ can be approximated in terms of $w$ by the functions $h_1(w)$ and $h_2(w)$ respectively as 

\begin{equation}\label{h1b}
    h_1(w)=a_2w^2+a_3w^3+O(w^4)
\end{equation}
    \begin{equation}\label{h2b}
    h_2(w)=b_2w^2+b_3w^3+O(w^4)
\end{equation}

The quasilinear partial differential equation, which the vector of functions
\begin{equation*}
    h=\begin{pmatrix}
      h_1\\
      h_2
    \end{pmatrix}
\end{equation*}

has to satisfy, is given by
\begin{equation}\label{a2dhw}
    Dh(w)[Aw+F(w,h(w))]-Bh(w)-g(w,h(w))=0
\end{equation}
where
\begin{equation}
g=\begin{pmatrix}
  g_1\\
  g_2
\end{pmatrix}
  ,  B = \begin{pmatrix}
      0 & 0\\
       0 & 0
    \end{pmatrix}
    ,~~F = f,~~A = \pm \sqrt{\frac{3}{2}}a
\end{equation}
Next we solve equation (\ref{a2dhw}) by substituting $A, h, F, B, g$ into it and equating equal powers of $w$ in order to obtain $h(w)$ up to the desired order. By comparing powers of $w$ from both sides of equation (\ref{a2dhw}) , we obtain the constants $a_2$, $a_3$, $b_2$ and $b_3$ as
    \begin{equation}
    a_2=\mp\frac{1}{2a},~~ a_3=0,~~b_2=0,~~b_3=0
\end{equation}
Finally the dynamics of the system restricted to the centre manifold is given by
\begin{equation}\label{a2dotw}
    \Dot{w}=Aw + F(w, h(w)) = \pm \sqrt{\frac{3}{2}}aw \mp \frac{1}{2a}\sqrt{\frac{3}{2}}w^3+O(w^4)
\end{equation}
At the lowest order of equation (\ref{a2dotw}) we have obtained an odd-parity term . Hence, like before, point  B for $n=3$ will be stable or unstable depending on the sign of the critical point.  \\

A similar analysis has been carried out for $n=4, 5, ....$, $a = any$. It has been found that in all the cases, the lowest order appears to be an odd-parity term with co-efficient $\pm \sqrt{\frac{3}{2}}a$ and thus the stability will depend upon the sign infront of the parameter $a$.

\subsection {For $n=2$, $a \ne 0, ~1$ }\label{appendixA4}
In this case also, the system of equations are not in the standard form. These system of variables can not be brought into the required form by change of variables. In such a situation, in CMT theory, the process is to find the eigen vectors corresponding to the Jacobian matrix and construct a matrix $S$ which is the matrix of the eigen vectors. In this particular case, for the Jacobian $J_3$, we find out the eigen vectors and construct the matrix $S$ which comes out as
$$S= \begin{pmatrix}
      0 ~~& 1 ~~& 0\\
       1 ~~& 0~~ & 0\\
       0 ~~& 0~~ & 0
    \end{pmatrix}$$
    The next step in CMT analysis is to diagonalize $S$. However, in this particular case, the  matrix $S$ corresponding to the Jacobian $J_3$ can not be diagonalized since $S$ turns out to be a singular matrix. Thus CMT theory fails for $n=2$, $a \ne 0,1$ case. \\
   
\section{For point C }\label{appendixB}
For the critical point C ($0, 1, 0$), in order to apply the
centre manifold theory, we first transform the system of equations (\ref{xprimezfinal})-(\ref{zprimezfinal}) to the standard form by rescaling the co-ordinates as 
\begin{equation}
X=x
\end{equation}
\begin{equation}
Y=y-1
\end{equation}
\begin{equation}
Z=z
\end{equation}
This rescaling moves the point (0, 1, 0) to the origin (0, 0, 0) of the
phase space.
Then equations (\ref{xprimezfinal})-(\ref{zprimezfinal}) becomes
\begin{equation}\label{xprimepointb}
    X'=\frac{1}{2}(1-Z)^{n-1} \{3X^3-3X(Y^2+2Y+2)\}-\sqrt{\frac{3}{2}}(Y^2+2Y+1)Z^{n-1}
\end{equation}
\begin{equation}\label{yprimepointb}
    Y'=-\frac{1}{2}(Y+1)(1-Z)^{n-1} \{-3X^2+3(Y^2+2Y)\}+\sqrt{\frac{3}{2}}X(Y+1)Z^{n-1}
\end{equation}
\begin{equation}\label{zprimepointb}
    Z'=\sqrt{6}X(n\alpha)^\frac{1}{n-1} (1-Z)^{n+1}
\end{equation}
The Jacobian matrix of this system of equations evaluated at $(0,0,0)$ is obtained as 
\begin{equation}
   J_5=    \begin{pmatrix}
      -3~~ & 0~~ & 0\\
       0~~ & -3~~ &0 \\
       \sqrt{6}(n\alpha)^\frac{1}{n-1}~~ & 0~~ & 0
    \end{pmatrix}      ~~~~~\mathrm{for}~~  n\ne 2
  \end{equation}
  As mentioned in section \ref{CMT}, the standard form of system of equations should not contain any term linear in other variables. This can be ensured by diagonalizing the system of equations. The system of equations can be brought into the diagonal form by introducing another set of new variables as
\begin{equation}\label{criticalb}
  \begin{pmatrix}
  u\\
  v\\
  w\\
    \end{pmatrix}  
    =
    \begin{pmatrix}
    0~~ & 1~~& 0\\
    
   - \sqrt{\frac{2}{3}}(n\alpha)^\frac{1}{n-1}~~ &0~~ &0\\
    \sqrt{\frac{2}{3}}(n\alpha)^\frac{1}{n-1}~~ &0~~ &1\\

    \end{pmatrix}
    \begin{pmatrix}
  X\\
  Y\\
  Z\\
    \end{pmatrix}  
    \end{equation}
 In the following sub-sections, we will show the detailed calculations for two values of $n$, viz, $n=3$ and $n=4$.    
\subsection {For $n=3$ }\label{appendixB1}
For $n = 3$, the transformed system (\ref{criticalb}) will take the form 
 \begin{equation}
  \begin{pmatrix}
  u\\
  v\\
  w\\
    \end{pmatrix}  
    =
    \begin{pmatrix}
    0~~ & 1~~& 0\\
    
   - \sqrt{2\alpha}~~ &0~~ &0\\
    \sqrt{2\alpha}~~ &0~~ &1\\

    \end{pmatrix}
    \begin{pmatrix}
  X\\
  Y\\
  Z\\
    \end{pmatrix}  
    \end{equation}
  which can also be written as 
  \begin{equation}
u = Y ~~\Rightarrow ~~Y = u
\end{equation}
\begin{equation}
v = - \sqrt{2\alpha}X ~~\Rightarrow ~~X = -\frac{1}{\sqrt{2\alpha}}v
\end{equation}
\begin{equation}
w = \sqrt{2\alpha}X + Z ~~\Rightarrow~~ Z = w + v
\end{equation}
 After these set of substitutions, equations (\ref{xprimepointb}) - (\ref{zprimepointb}) take the final form 
 \begin{eqnarray}
u' = -3u + g_1(u,v,w) \\\nonumber
v' = -3 v + g_2(u,v,w) \\\nonumber
w' = 0 + f(u,v,w)
\end{eqnarray}
where $g_1, g_2$ and $f$ are polynomials in $(u, v, w)$ of degree greater than 2 given by
\begin{multline}
         g_1=- \frac{1}{2}(-\frac{3v^2}{2\alpha}+3u^2)-\frac{1}{2}u[-\frac{3v^2}{2\alpha}+3(u^2+2u)]\\-\frac{1}{2}(u+1)[(w+v)^2-2(w+v)][-\frac{3v^2}{2\alpha}+3(u^2+2u)]
         -\frac{1}{2}\sqrt{\frac{3}{\alpha}}v(w+v)^2(u+1) \nonumber
    \end{multline}

 \begin{multline}
         g_2=- \frac{1}{2}[-\frac{3v^3}{2\alpha}+3v(u^2+2u)]-\frac{1}{2}[(w+v)^2-2(w+v)][-\frac{3v^3}{2\alpha}+3v(u^2+2u+2)]\\
         +\sqrt{3\alpha}(w+v)^2(u^2+2u+1)
        \nonumber
    \end{multline}

\begin{multline}
         f= \frac{1}{2}\lbrace-\frac{3v^3}{2\alpha}+3v(u^2+2u)\rbrace+\frac{1}{2}\lbrace(w+v)^2-2(w+v)\rbrace\lbrace-\frac{3v^3}{2\alpha}+3v(u^2+2u+2)\rbrace\\
         -\sqrt{3\alpha}(w+v)^2(u^2+2u+1)-3v[(w+v)^2-2(w+v)]-3v[(w+v)^2-2(w+v)][1-2(w+v)+(w+v)^2] \nonumber
    \end{multline}

In terms of these new set of variables, the system of equations can also be represented in matrix form as
   \begin{equation}
  \begin{pmatrix}

  u'\\
  v'\\
  w'\\
    \end{pmatrix}  
    =
    \begin{pmatrix}
   -3 & 0& 0\\
    
    0 &-3 &0\\
    0& 0& 0&\\

    \end{pmatrix}
    \begin{pmatrix}
  u\\
  v\\
  w\\
    \end{pmatrix}  
   +
\begin{pmatrix}
  g_1\\
  g_2\\
  f\\
    \end{pmatrix} 
    \end{equation}
Now the coordinates which correspond to non-zero eigenvalues $(u, v)$ can be approximated in terms of $w$ by the functions $h_1(w)$ and $h_2(w)$ respectively as 

\begin{equation}\label{h1b}
    h_1(w)=a_2w^2+a_3w^3+O(w^4)
\end{equation}
    \begin{equation}\label{h2b}
    h_2(w)=b_2w^2+b_3w^3+O(w^4)
\end{equation}

The quasilinear partial differential equation, which the vector of functions
\begin{equation*}
    h=\begin{pmatrix}
      h_1\\
      h_2
    \end{pmatrix}
\end{equation*}

has to satisfy, is given by
\begin{equation}\label{bdhw}
    Dh(w)[Aw+F(w,h(w))]-Bh(w)-g(w,h(w))=0
\end{equation}
where
\begin{equation}
g=\begin{pmatrix}
  g_1\\
  g_2
\end{pmatrix}
  ,  B = \begin{pmatrix}
      -3 & 0\\
       0 & -3
    \end{pmatrix}
    ,~~F = f,~~A = 0
\end{equation}
Next we solve equation (\ref{bdhw}) by substituting $A, h, F, B, g$ into it and equating equal powers of $w$ in order to obtain $h(w)$ up to the desired order. By comparing powers of $w$ from both sides of equation (\ref{bdhw}) for $h_1(w)$ and $h_2(w)$, we obtain the constants $a_2$, $a_3$, $b_2$ and $b_3$ as
    \begin{equation}
    a_2=0,~~ a_3=0,~~b_2=\sqrt{\frac{\alpha}{3}},~~b_3=-\frac{\sqrt{\frac{\alpha}{3}}}{3}(6+4\sqrt{3\alpha})
\end{equation}
Finally the dynamics of the system restricted to the centre manifold is given by
\begin{equation}\label{bdotw}
    \Dot{w}=Aw + F(w, h(w)) = -\sqrt{3\alpha}w^2+2(\sqrt{3\alpha} - \alpha)w^3+O(w^4)
\end{equation}
At the lowest order of equation (\ref{bdotw}) we have obtained an even-parity term. Hence according to CMT theory, for $n=3$ point  C will always be unstable.  
\subsection {n=4 }\label{appendixB2}
In this case, the transformed system (\ref{criticalb}) will take the form 
 \begin{equation}
  \begin{pmatrix}
  u\\
  v\\
  w\\
    \end{pmatrix}  
    =
    \begin{pmatrix}
    0~~ & 1~~& 0\\
    
   - \sqrt{\frac{2}{3}}(4\alpha)^\frac{1}{3}~~ &0~~ &0\\
  \sqrt{\frac{2}{3}}(4\alpha)^\frac{1}{3}~~ &0~~ &1\\

    \end{pmatrix}
    \begin{pmatrix}
  X\\
  Y\\
  Z\\
    \end{pmatrix}  
    \end{equation}
  which can also be written as 
  \begin{equation}
u = Y ~~\Rightarrow ~~Y = u
\end{equation}
\begin{equation}
v = - mX ~~\Rightarrow ~~X = -\frac{v}{m}
\end{equation}
\begin{equation}
w = mX + Z ~~\Rightarrow~~ Z = w + v
\end{equation}
where $ m=\sqrt{\frac{2}{3}}(4\alpha)^\frac{1}{3}$\\
 
 After these set of substitutions, equations (\ref{xprimepointb}) - (\ref{zprimepointb}) take the final form 
 \begin{eqnarray}
u' = -3u + g_3(u,v,w) \\\nonumber
v' = -3 v + g_4(u,v,w) \\\nonumber
w' = 0 + f1(u,v,w)
\end{eqnarray}
with the polynomials $g_1, g_2$ and $f$ given by
\begin{multline}
         g_3=- \frac{1}{2}(-\frac{3v^2}{m^2}+3u^2)-\frac{1}{2}u[-\frac{3v^2}{m^2}+3(u^2+2u)]\\-\frac{1}{2}(u+1)[3(w+v)^2-3(w+v)-(w+v)^3][-\frac{3v^2}{m^2}+3(u^2+2u)]
         -\sqrt{\frac{3}{2}}\frac{v}{m}(u+1)(w+v)^3 \nonumber
    \end{multline}

 \begin{multline}
         g_4=- \frac{1}{2}[-\frac{3v^3}{m^2}+3v(u^2+2u)]-\frac{1}{2}[3(w+v)^2-3(w+v)-(w+v)^3][-\frac{3v^3}{m^2}+3v(u^2+2u+2)]\\
         +\sqrt{\frac{3}{2}}m(w+v)^3(u^2+2u+1)
        \nonumber
    \end{multline}

\begin{multline}
         f_1= \frac{1}{2}[-\frac{3v^3}{m^2}+3v(u^2+2u)]+\frac{1}{2}[3(w+v)^2-3(w+v)-(w+v)^3][-\frac{3v^3}{m^2}+3v(u^2+2u+2)]\\
         -\sqrt{\frac{3}{2}}m(w+v)^3(u^2+2u+1)-\\3v[-3(w+v)+3(w+v)^2-(w+v)^3]-3v[(w+v)^2-2(w+v)][1-3(w+v)+3(w+v)^2-(w+v)^3] \nonumber
    \end{multline}
      
 In terms of these new set of variables, the system of equations can also be represented in matrix form as
   \begin{equation}
  \begin{pmatrix}

  u'\\
  v'\\
  w'\\
    \end{pmatrix}  
    =
    \begin{pmatrix}
   -3 & 0& 0\\
    
    0 &-3 &0\\
    0& 0& 0&\\

    \end{pmatrix}
    \begin{pmatrix}
  u\\
  v\\
  w\\
    \end{pmatrix}  
   +
\begin{pmatrix}
  g_3\\
  g_4\\
  f_1\\
    \end{pmatrix} 
    \end{equation}
Now the coordinates which correspond to non-zero eigenvalues $(u, v)$ can be approximated in terms of $w$ by the functions $h_1(w)$ and $h_2(w)$ respectively as 

\begin{equation}\label{h1}
    h_1(w)=a_2w^2+a_3w^3+O(w^4)
\end{equation}
    \begin{equation}\label{h2}
    h_2(w)=b_2w^2+b_3w^3+O(w^4)
\end{equation}

The quasilinear partial differential equation, which the vector of functions
\begin{equation*}
    h=\begin{pmatrix}
      h_1\\
      h_2
    \end{pmatrix}
\end{equation*}

has to satisfy, is given by
\begin{equation}\label{cdhw1}
    Dh(w)[Aw+F(w,h(w))]-Bh(w)-g(w,h(w))=0
\end{equation}
where
\begin{equation}
g=\begin{pmatrix}
  g_3\\
  g_4
\end{pmatrix}
  ,  ~B = \begin{pmatrix}
      -3 & 0\\
       0 & -3
    \end{pmatrix}
    ,~~F = f_1,~~A = 0
\end{equation}
Next we solve equation (\ref{cdhw1}) by substituting $A, h, F, B, g$ into it and equating equal powers of $w$ in order to obtain $h(w)$ up to the desired order. By comparing powers of $w$ as before, we obtain the constants $a_2$, $a_3$, $b_2$ and $b_3$ as
    \begin{equation}
    a_2=0,~~ a_3=0,~~b_2=0,~~b_3=\frac{1}{3}\sqrt{\frac{3}{2}}m
\end{equation}
Finally the dynamics of the system restricted to the centre manifold is given by
\begin{equation}\label{cdotw4}
    \Dot{w}=Aw + F(w, h(w)) = -\sqrt{\frac{3}{2}}m w^3+O(w^4)
\end{equation}
At the lowest order of equation (\ref{cdotw4}) we have obtained an odd-parity term with negative coefficient. Hence according to CMT theory, point  C will always be stable for $n=4$.  
\section{Detailed analysis for point D }\label{appendixC}
For the critical point D ($0, -1, 0$), as before we transform the system of equations (\ref{xprimezfinal})-(\ref{zprimezfinal}) to the standard form by  the following co-ordinates rescaling :  
\begin{equation}
X=x
\end{equation}
\begin{equation}
Y=y+1
\end{equation}
\begin{equation}
Z=z
\end{equation}
This rescaling moves the point (0, -1, 0) to the origin (0, 0, 0) of the
phase space. Then equations (\ref{xprimezfinal})-(\ref{zprimezfinal}) becomes
\begin{equation}\label{xprimepointc}
X'=\frac{1}{2}(1-Z)^{n-1} \{3X^3-3X(Y^2-2Y+2)\}-\sqrt{\frac{3}{2}}(Y^2-2Y+1)Z^{n-1}
 \end{equation}
\begin{equation}\label{yprimepointc}
 Y'=-\frac{1}{2}(Y-1)(1-Z)^{n-1} \{-3X^2+3(Y^2-2Y)\}+\sqrt{\frac{3}{2}}X(Y-1)Z^{n-1}
\end{equation}
\begin{equation}\label{zprimepointc}
     Z'=\sqrt{6}X(n\alpha)^\frac{1}{n-1} (1-Z)^{n+1}
\end{equation}
The Jacobian matrix for this system of equations evaluated at $(0,0,0)$ is obtained as 
\begin{equation}
   J_5=    \begin{pmatrix}
      -3~~ & 0~~ & 0\\
       0~~ & -3~~ &0 \\
       \sqrt{6}(n\alpha)^\frac{1}{n-1}~~ & 0~~ & 0
    \end{pmatrix}      ~~~~~\mathrm{for}~~  n\ne 2
  \end{equation}
  The system of equations are then brought into the diagonal form by introducing a new set of variables 
\begin{equation}\label{criticalc}
  \begin{pmatrix}
  u\\
  v\\
  w\\
    \end{pmatrix}  
    =
    \begin{pmatrix}
    0~~ & 1~~& 0\\
    
   - \sqrt{\frac{2}{3}}(n\alpha)^\frac{1}{n-1}~~ &0~~ &0\\
    \sqrt{\frac{2}{3}}(n\alpha)^\frac{1}{n-1}~~ &0~~ &1\\

    \end{pmatrix}
    \begin{pmatrix}
  X\\
  Y\\
  Z\\
    \end{pmatrix}  
    \end{equation}
 In the following sub-sections, like before, we will show the detailed calculations for $n=3$ and $n=4$.    
 \subsection {For $n=3$ }\label{appendixC1}
For $n = 3$, the transformed system (\ref{criticalc}) takes the form 
 \begin{equation}
  \begin{pmatrix}
  u\\
  v\\
  w\\
    \end{pmatrix}  
    =
    \begin{pmatrix}
    0~~ & 1~~& 0\\
    
   - \sqrt{2\alpha}~~ &0~~ &0\\
    \sqrt{2\alpha}~~ &0~~ &1\\

    \end{pmatrix}
    \begin{pmatrix}
  X\\
  Y\\
  Z\\
    \end{pmatrix}  
    \end{equation}
  which can be rewritten as 
  \begin{equation}
u = Y ~~\Rightarrow ~~Y = u
\end{equation}
\begin{equation}
v = - \sqrt{2\alpha}X ~~\Rightarrow ~~X = -\frac{1}{\sqrt{2\alpha}}v
\end{equation}
\begin{equation}
w = \sqrt{2\alpha}X + Z ~~\Rightarrow~~ Z = w + v
\end{equation}
 After these set of substitutions, equations (\ref{xprimepointc}) - (\ref{zprimepointc}) take the final form 
 \begin{eqnarray}
u' = -3u + g_1(u,v,w) \\\nonumber
v' = -3 v + g_2(u,v,w) \\\nonumber
w' = 0 + f(u,v,w)
\end{eqnarray}
where $g_1, g_2$ and $f$ are polynomials in $(u, v, w)$ given by
\begin{multline}
         g_1= \frac{1}{2}(-\frac{3v^2}{2\alpha}+3u^2)-\frac{1}{2}u[-\frac{3v^2}{2\alpha}+3(u^2-2u)]\\-\frac{1}{2}(u-1)[(w+v)^2-(w+v)][-\frac{3v^2}{2\alpha}+3(u^2-2u)]
         -\frac{1}{2}\sqrt{\frac{3}{\alpha}}v(w+v)^2(u-1) \nonumber
    \end{multline}

 \begin{multline}
         g_2=- \frac{1}{2}[-\frac{3v^3}{2\alpha}+3v(u^2-2u)]-\frac{1}{2}[(w+v)^2-2(w+v)][-\frac{3v^3}{2\alpha}+3v(u^2-2u+2)]\\
         +\sqrt{3\alpha}(w+v)^2(u^2-2u+1)
        \nonumber
    \end{multline}

\begin{multline}
         f= \frac{1}{2}\lbrace-\frac{3v^3}{2\alpha}+3v(u^2-2u)\rbrace+\frac{1}{2}\lbrace(w+v)^2-2(w+v)\rbrace\lbrace-\frac{3v^3}{2\alpha}+3v(u^2-2u+2)\rbrace\\
         -\sqrt{3\alpha}(w+v)^2(u^2-2u+1)-3v[(w+v)^2-2(w+v)]-3v[(w+v)^2-2(w+v)][1-2(w+v)+(w+v)^2] \nonumber
    \end{multline}

The system of equations in matrix form becomes
   \begin{equation}
  \begin{pmatrix}

  u'\\
  v'\\
  w'\\
    \end{pmatrix}  
    =
    \begin{pmatrix}
   -3 & 0& 0\\
    
    0 &-3 &0\\
    0& 0& 0&\\

    \end{pmatrix}
    \begin{pmatrix}
  u\\
  v\\
  w\\
    \end{pmatrix}  
   +
\begin{pmatrix}
  g_1\\
  g_2\\
  f\\
    \end{pmatrix} 
    \end{equation}
Proceeding in a similar manner, by solving the differential equation  
\begin{equation}
    Dh(w)[Aw+F(w,h(w))]-Bh(w)-g(w,h(w))=0 \nonumber
\end{equation}
for $h_1(w)$ and $h_2(w)$ with 
\begin{equation}
g=\begin{pmatrix}
  g_1\\
  g_2
\end{pmatrix}
  ,  B = \begin{pmatrix}
      -3 & 0\\
       0 & -3
    \end{pmatrix}
    ,~~F = f,~~A = 0 \nonumber
\end{equation}
and equating equal powers of $w$, we obtain the constants $a_2$, $a_3$, $b_2$ and $b_3$ as
    \begin{equation}
    a_2=0,~~ a_3=0,~~b_2=\frac{\sqrt{3\alpha}}{3},~~b_3=\frac{\sqrt{3\alpha}}{9}(6+4\sqrt{3\alpha})
\end{equation}
Finally the dynamics of the system is given by
\begin{equation}\label{cdotw2}
    \Dot{w}=Aw + F(w, h(w)) = -\sqrt{3\alpha}w^2+(6-2\sqrt{3}\alpha)w^3 + O(w^4)
\end{equation}
At the lowest order of equation (\ref{cdotw2}) we again obtain an even-parity term which indicates that for $n=3$ the dynamics of the system around point D will always be unstable.
\subsection {For $n=4$ }\label{appendixC2}
For $n = 4$, the transformed system (\ref{criticalc}) takes the form 
 \begin{equation}
  \begin{pmatrix}
  u\\
  v\\
  w\\
    \end{pmatrix}  
    =
    \begin{pmatrix}
    0~~ & 1~~& 0\\
    
   - \sqrt{\frac{2}{3}}(4\alpha)^\frac{1}{3}~~ &0~~ &0\\
  \sqrt{\frac{2}{3}}(4\alpha)^\frac{1}{3}~~ &0~~ &1\\

    \end{pmatrix}
    \begin{pmatrix}
  X\\
  Y\\
  Z\\
    \end{pmatrix}  
    \end{equation}
  which can also be written as 
  \begin{equation}
u = Y ~~\Rightarrow ~~Y = u
\end{equation}
\begin{equation}
v = - mX ~~\Rightarrow ~~X = -\frac{v}{m}
\end{equation}
\begin{equation}
w = mX + Z ~~\Rightarrow~~ Z = w + v
\end{equation}
where $ m=\sqrt{\frac{2}{3}}(4\alpha)^\frac{1}{3}$\\
 
 After these set of substitutions, equations (\ref{xprimepointc}) - (\ref{zprimepointc}) take the final form 
 \begin{eqnarray}
u' = -3u + g_1(u,v,w) \\\nonumber
v' = -3 v + g_2(u,v,w) \\\nonumber
w' = 0 + f(u,v,w)
\end{eqnarray}
where $g_1, g_2$ and $f$ are polynomials in $(u, v, w)$ of degree greater than 2 given by
\begin{multline}
         g_1= \frac{1}{2}(-\frac{3v^2}{m^2}+3u^2)-\frac{1}{2}u[-\frac{3v^2}{m^2}+3(u^2-2u)]\\-\frac{1}{2}(u-1)[3(w+v)^2-3(w+v)-(w+v)^3][-\frac{3v^2}{m^2}+3(u^2-2u)]
         -\sqrt{\frac{3}{2}}\frac{v}{m}(u-1)(w+v)^3 \nonumber
    \end{multline}

 \begin{multline}
         g_2=- \frac{1}{2}[-\frac{3v^3}{m^2}+3v(u^2-2u)]-\frac{1}{2}[3(w+v)^2-3(w+v)-(w+v)^3][-\frac{3v^3}{m^2}+3v(u^2-2u+2)]\\
         +\sqrt{\frac{3}{2}}m(w+v)^3(u^2-2u+1)
        \nonumber
    \end{multline}

\begin{multline}
         f= \frac{1}{2}[-\frac{3v^3}{m^2}+3v(u^2-2u)]+\frac{1}{2}[3(w+v)^2-3(w+v)-(w+v)^3][-\frac{3v^3}{m^2}+3v(u^2-2u+2)]\\
         -\sqrt{\frac{3}{2}}m(w+v)^3(u^2-2u+1)-3v[-3(w+v)+3(w+v)^2-(w+v)^3]-3v[(w+v)^2-2(w+v)][1-3(w+v)+3(w+v)^2-(w+v)^3] \nonumber
    \end{multline}
      
  The system of equations can then be represented in matrix form as
   \begin{equation}
  \begin{pmatrix}

  u'\\
  v'\\
  w'\\
    \end{pmatrix}  
    =
    \begin{pmatrix}
   -3 & 0& 0\\
    
    0 &-3 &0\\
    0& 0& 0&\\

    \end{pmatrix}
    \begin{pmatrix}
  u\\
  v\\
  w\\
    \end{pmatrix}  
   +
\begin{pmatrix}
  g_1\\
  g_2\\
  f\\
    \end{pmatrix} 
    \end{equation}
Again by following the same procedure, the differential equation 
\begin{equation}\label{cdhw}
    Dh(w)[Aw+F(w,h(w))]-Bh(w)-g(w,h(w))=0 \nonumber
\end{equation}
is solved for $h_1(w)$ and $h_2(w)$ with  
\begin{equation}
g=\begin{pmatrix}
  g_1\\
  g_2
\end{pmatrix}
  ,  B = \begin{pmatrix}
      -3 & 0\\
       0 & -3
    \end{pmatrix}
    ,~~F = f,~~A = 0
\end{equation}
Next by comparing powers of $w$,  we obtain the constants $a_2$, $a_3$, $b_2$ and $b_3$ as
    \begin{equation}
    a_2=0,~~ a_3=0,~~b_2=0,~~b_3=\frac{1}{3}\sqrt{\frac{3}{2}}m
\end{equation}
Finally the dynamics of the system restricted to the centre manifold is given by
\begin{equation}\label{cdotw}
    \Dot{w}=Aw + F(w, h(w)) = -(4\alpha )^\frac{1}{3}w^3+O(w^4)
\end{equation}
At the lowest order of equation (\ref{cdotw}) we have obtained an odd-parity term with negative coefficient. Hence according to CMT theory, for $n=4$ point D will always be stable. 

\end{document}